\documentclass[aps,prb,twocolumn,superscriptaddress,amsmath,amssymb,floatfix]{revtex4-1}

\usepackage{graphicx}
\usepackage{dcolumn}
\usepackage{bm}
\usepackage{natbib}
\usepackage{verbatim}
\usepackage{mathptmx}
\usepackage{color}
\usepackage[T1]{fontenc}

\begin{document}

\widetext

\title{Fate of the open-shell singlet ground state in the experimentally accessible acenes: a quantum Monte Carlo study}

\begin{abstract}
By means of the Jastrow correlated antisymmetrized geminal power (JAGP) wave function and quantum Monte Carlo (QMC) methods, we study the ground state properties of the oligoacene series, up to the nonacene. The JAGP is the accurate variational realization of the resonating-valence-bond (RVB) ansatz proposed by Pauling and Wheland to describe aromatic compounds. We show that the long-ranged RVB correlations built in the acenes' ground state are detrimental for the occurrence of open-shell diradical or polyradical instabilities, previously found by lower-level theories. We substantiate our outcome by a direct comparison with another wave function, tailored to be an open-shell singlet (OSS) for long-enough acenes. By comparing on the same footing the RVB and OSS wave functions, both optimized at a variational QMC level, and further projected by the lattice regularized diffusion Monte Carlo (LRDMC) method, we prove that the RVB  wave function has always a lower variational energy and better nodes than the OSS, for all molecular species considered in this work. The entangled multi-reference RVB state 
acts against the electron edge localization implied by the OSS wave function, and weakens the diradical tendency for higher oligoacenes. These properties are reflected by several descriptors, including wave function parameters, bond length alternation, aromatic indices, and spin-spin correlation functions. In this context, we propose a new aromatic index estimator suitable for geminal wave functions. For the largest acenes taken into account, the long-range decay of the charge-charge correlation functions is compatible with a quasi-metallic behavior.
\end{abstract}

\author{Nicolas Dupuy}
\affiliation{Institut de Min\'eralogie, de Physique des Mat\'eriaux et de Cosmochimie (IMPMC), Sorbonne Universit\'e, CNRS UMR 7590, IRD UMR 206, MNHN, 4 Place Jussieu, 75252 Paris, France}
\author{Michele Casula}
\email{michele.casula@upmc.fr}
\affiliation{Institut de Min\'eralogie, de Physique des Mat\'eriaux et de Cosmochimie (IMPMC), Sorbonne Universit\'e, CNRS UMR 7590, IRD UMR 206, MNHN, 4 Place Jussieu, 75252 Paris, France}

\date{\today}

\maketitle

\section{Introduction}
\label{acenes:intro}

Zig-zag graphene nanoribbons have attracted a great deal of interest for the possibility of building ribbon-based spintronic devices. Indeed, quite recent band structure calculations found edge states located at the Fermi level\cite{nakada1996}. In spin-symmetry broken DFT solutions, a gap opens and the zig-zag edges become spin polarized and antiferromagnetically ordered. According to \emph{ab initio} calculations\cite{nature_louie}, the application of an in-plane electric field perpendicular to the graphene ribbon could lead to a half-metal, where only one spin species carries electric current. So far, this proposal still awaits experimental confirmation. The main bottleneck is the synthesis of such systems, due to the extreme reactivity of their zig-zag edges that makes them unstable, also from a thermodynamical point of view\cite{wassmann2008}. However, significant progress has been made in the last couple of years towards their generation, through controlled chemical reaction cascades\cite{dorel2017}, and properly chosen molecular precursors\cite{ruffieux2015}.

Polyacenes are the narrowest zig-zag graphene nanoribbons. In this work, we will study their finite-length version, i.e. the oligoacenes. We will take into account various sizes, going from the anthracene (3 fused benzene rings) to the nonacene (9 rings). While sharing some important similarities with their infinite size counterpart, they offer several advantages. The most important one is that they exist in nature in stable forms, although their stability decreases rapidly with their length. Despite this fact, synthesis of oligoacenes with up to 9 rings has been reported. Seven, eight, and nine fused rings\cite{zade_heptacene_2010,tonshoff2010photogeneration} are created in inert matrices\cite{mondal_synthesis_2009}, which protect and stabilize them. 
Thus, a direct comparison between some theoretical and experimental properties can be made up to the nonacene. 

The second reason to study the oligoacenes is that their finite size is small enough to attempt the solution of their electronic structure by high-level theory methods\cite{chan_dmrg,hajgato2009benchmark,plasser_multiradical_2013,rivero2013,ibeji2015,yang2016,fosso2016,battaglia2017,lee2017}, including quantum Monte Carlo. 
Looking for their ground state (GS) is a challenging task, and applying advanced but more expensive numerical methods is necessary, because these systems belong - with no doubts - to the strongly correlated family. Indeed, the gap between their highest occupied molecular orbital (HOMO) and the lowest unoccupied one (LUMO) closes as a function of the polymer length. Moreover, the density of states around the HOMO-LUMO gap gets higher and higher as the length increases. Therefore, static correlation effects become relevant already at moderate length. On top of that, charge and spin fluctuations play a crucial role, because the low-energy physics is driven by $\pi$ orbitals accommodating one electron per carbon site, arranged in the $sp^2$ ($\sigma$) network. This half-filled situation is the most correlated one. It can be modeled in the framework of the resonating-valence-bond (RVB) theory,
where the ground state is a linear combination of an exponentially growing number of configurations, made by all possible singlet bonds connecting two carbon sites, and by avoiding double occupancy on $p_z$ orbitals. Clar's theory for benzenoid rings\cite{clar_polycyclic_1964}, widely applied to polycyclic aromatic hydrocarbons (PAHs), is the chemical graph theory 
analogue of the RVB ansatz\cite{misra2009}.

Last but not least, oligoacenes have been proposed as fundamental bricks of organic electronics\cite{bendikov2004,anthony_larger_2008,ye2014}, in a large variety of possible applications\cite{pilevarshahri_spin_2011,hardin2012,yelin_conductance_2016}. One of the latest proposals, which stimulated a great amount of research on their GS and excited states properties, is the possibility of using them as organic photovoltaics. They would possess enhanced power-conversion efficiency thanks to the excited-state singlet fission into a couple of more stable triplet energy levels, without fluorescence decay\cite{zimmerman_singlet_2010,zimmerman_mechanism_2011}. Therefore, oligoacenes are promising for near-future technological applications.  

While there is an agreement on the closed-shell singlet (CSS) character of the smallest molecules GS (anthracene, tetracene, and pentacene), in literature there is no clear consensus on longer $n$-acenes, yet. Their GS nature has been highly debated along the years. A spin triplet instability has been first proposed for $n > 8$, based on spectroscopical and theoretical data\cite{angliker_electronic_1982,houk2001}, the rationale behind this being the tiny HOMO-LUMO gap. Later, unrestricted B3LYP calculations yielded an open-shell singlet GS solution\cite{bendikov2004oligoacenes}, starting from $n=6$. This state is diradical, namely it has two singly occupied molecular orbitals, each one localized on one of the two long edges. As the GS is singlet, each side is spin polarized with opposite spin orientation per side. These electronic properties can be 
related
to the antiferromagnetism found in wider zig-zag graphene nanoribbons. Bendikov and coworkers associated their diradical character with low chemical stability \cite{bendikov2004oligoacenes,zade2012reactivity}. This proposal triggered a significant amount of theoretical\cite{chan_dmrg,jiang2008electronic,hajgato2009benchmark,ess2010,hajgato_focal_2011,plasser_multiradical_2013,rivero2013,ibeji2015,yang2016,fosso2016,battaglia2017,lee2017,chai2017} and experimental studies\cite{mondal_synthesis_2009,zade_heptacene_2010,tonshoff2010photogeneration,purushothaman2011,zuzak2017,dorel2017}, aiming at validating the open-shell singlet (OSS) nature of the large acenes GS. The challenge is to capture the behavior of the GS as a function of the oligoacene length, being the electronic correlation stronger and stronger as the size increases. OSS has been found by spin-polarized GGA\cite{jiang2008electronic}, and by DMRG (CASCI) calculations in a STO-3G basis set\cite{chan_dmrg}, where the $\pi$ electrons are explicitly correlated. The latter two studies claimed that higher acenes are polyradical. Indeed, they found that there is more than one spin accumulating on each side, and the corresponding natural orbital occupations turn out to be fractional not only for the highest occupied natural orbital (HONO) and the lowest unoccupied one (LUNO), but also for those lying just below, or just above, their actual occupation being related to the size of the molecule. While the CASCI is potentially more accurate than the B3LYP and GGA functionals, it suffers however for its high computational cost and is limited to 
small basis sets. Very recent CASSCF\cite{battaglia2017} and particle-particle RPA\cite{yang2016} calculations gave the same qualitative answer, however they significantly reduced the impact of diradicality or polyradicality for small molecules, and pushed the CSS-to-OSS crossover to larger acenes. These latest studies suggest that one can start talking about diradicality only after 10 fused rings, while from $n=6$ to $n=9$ the natural orbital occupations depart significantly from the CSS. Obvious limits of these calculations are the size of the active space for CASSCF, and the quality of the particle-particle RPA approximation. In contrast to these results, very accurate single-reference coupled cluster (CC) calculations performed with focal point analysis, namely with extrapolation in both theory and basis set, give a robust closed-shell singlet as the true ground state, at least up to the undecacene ($n=11$)\cite{hajgato_focal_2011}. However, this approach could miss important static correlation, because it neglects multi-reference states, although the T1 single-excitation diagnostic is in its favor. More recent multi-reference CC calculations at average quadratic level of theory (MR-AQCC) highlighted that one should take into account not only the T1, but also the D2 double-excitation diagnostic, which instead suggests a breakdown of the acceptable single-reference threshold, for large enough acenes\cite{plasser_multiradical_2013}. On the other hand, the inclusion of dynamic correlations in the 
$\sigma$-$\pi$ excitation channels have shown to drastically reduce the diradical character at the CAS valence bond (CASVB) level\cite{lee2017}.

Given the very controversial nature of such an important class of systems, which might be relevant for spintronic, electronic and photonic applications at both molecular and solid state levels, it is highly desirable to have an alternative approach to settle the problem of their ground state characterization. We are going to use quantum Monte Carlo methods, which are very accurate in this situation. Indeed, we will use the Jastrow correlated antisymmetric geminal power (JAGP)\cite{michele_agp0,Casula2004}, a correlated variational wave function that is the \emph{ab initio} realization of the RVB or Clar's ansatz, particularly suited to describe PAHs. After optimizing this wave function at the variational Monte Carlo (VMC) level, we are going to further project the RVB wave function to the ground state of the system by lattice regularized diffusion Monte Carlo (LRDMC) simulations, in the fixed-node approximation. Within the flexible JAGP ansatz, we can select different classes of solutions, i.e. the spin triplet, CSS, OSS, and the fully resonating valence bond with long-range singlets (full JAGP or RVB), in order to compare their variational energies, and infer the true ground state nature of the acenes up to nine rings.

In Sec.~\ref{acenes:met}, we will detail the analytic form of the wave functions and their related properties, in Sec.~\ref{acenes:results} we will discuss the results, and finally in Sec.~\ref{acenes:conclu} we will draw the conclusions and perspectives.

\section{Methods}
\label{acenes:met}

\subsection{Wave functions}
\label{acenes:wfs}

\subsubsection{General form, basis set, pseudopotentials}

The QMC calculations have been carried out for a first-principles Hamiltonian with carbon core electrons replaced by a pseudopotential.
The carbon atom is described by a Hartree-Fock (HF) energy consistent 
pseudopotential with scalar relativistic corrections by Burkatzki \emph{et 
al.}\cite{filippi_pseudo}, while the Coulomb singularity of the hydrogen 
electron-ion potential has been replaced by a short-range 
non-diverging pseudopotential obtained within the same HF energy-consistent 
framework\cite{filippi_private}. The HF energy-consistent 
pseudopotentials are particularly suited for correlated quantum chemistry 
calculations. 

The electron correlation in the acenes family is described by wave functions $\Psi$ written as product of a Jastrow factor and a determinantal
part $\Delta$:
\begin{equation}
\Psi(\textbf{r})  = e^{-J(\textbf{r})} \cdot \Delta (\textbf{r}),
\label{acenes:eq:wave_function}
\end{equation}
where
$\textbf{r}=\{\textbf{r}^\uparrow_1,\ldots,\textbf{r}^\uparrow_{N^\uparrow},\textbf{r}^\downarrow_1,\ldots,\textbf{r}^\downarrow_{N^\downarrow}\}$
is the $N$-electron position, with the total number of electrons
$N=N^\uparrow+N^\downarrow$, the sum of the up- and down-spin components.

We are going to use three types of determinants $\Delta$, described
below, including static electron correlations at different levels,
while the dynamic correlations are taken into account by the Jastrow
factor, whose form is kept the same for all cases. 

The Jastrow factor exponent $J$ in Eq.~\ref{acenes:eq:wave_function} takes the form of $J_1+J_2+J_{3/4}$, where it
is decomposed into one- ($J_1$), two- ($J_2$), and three/four-body
($J_{3/4}$) terms. Our Jastrow function is spin-independent, therefore the
$N$-electron coordinates in its argument will be denoted as $(\textbf{r}_1,...,\textbf{r}_N)$, 
without specifying the spin of the particles. 

For the one-body $J_1$ part, we use the form
\begin{equation}
J_1(\textbf{r}_1,...,\textbf{r}_N) = \sum\limits_i^N \sum\limits_a^{N_\textrm{atoms}}\sum\limits_\mu^{N_\textrm{basis}} f_\mu^a ~\chi^J_{a,\mu} (\textbf{r}_i),
\label{acenes:eq:one_body}
\end{equation}
where $i$ runs over the $N$ electron coordinates, and $\mu$ runs over
the Jastrow basis-set functions $\chi^J_{a,\mu}(\mathbf{r}) \equiv \chi^J_{a,\mu}(\mathbf{r}-\mathbf{q}_a)$, centered on the $a$-th nucleus located at $\mathbf{q}_a$. Analogously to $\mathbf{r}$, we define the collective position of the $N_{\textrm{atoms}}$ atoms in the system as the vector $\mathbf{q}=\{\mathbf{q}_1,\ldots,
\mathbf{q}_{N_\textrm{atoms}}\}$. $N_\textrm{basis}$ is the basis set dimension, which in general depends on each atomic site $a$.

The two-body $J_2$ part reads
\begin{equation}
J_2 (\textbf{r}_1,...,\textbf{r}_N) = \sum\limits_{1\le i<j\le N} u(|\textbf{r}_i-\textbf{r}_j|),
\label{acenes:eq:two_body}
\end{equation}
where $u$ is the radial function 
\begin{equation} 
u(\emph{r})=\frac{1}{2\gamma}(1-e^{-\gamma\emph{r}}),
\label{acenes:eq:u_jas}
\end{equation}
with $\gamma$ a positive term to be optimized. The form in Eq.~\ref{acenes:eq:u_jas} fulfills the Kato cusp conditions\cite{kato} for unlike-spin particles, by removing the divergence of the local energy $H\Psi(\textbf{r})/\Psi(\textbf{r})$ at the electron coalescence points due to the singularity of the electron-electron Coulomb potential. 

Finally, the three- and four-body $J_{3/4}$ contributions are written as
\begin{equation}
J_{3/4}(\textbf{r}_1,...,\textbf{r}_N) = 
\sum\limits_{i,j}^N
\sum\limits_{a,b}^{N_\textrm{atoms}}
\sum\limits_{\mu,\nu}^{N_\textrm{basis}}
g_{\mu,\nu}^{a,b}~\chi^J_{a,\mu}(\textbf{r}_i)\chi^J_{b,\nu}(\textbf{r}_j)
\label{acenes:eq:34_body}
\end{equation}
where $\chi^J_{a,\mu}$ belong to the same atomic basis set as for $J_1$. Therefore, $J_{3/4}$ correlates two electrons ($i$ and $j$) under the influence of one nucleus (in case $a=b$) or
two nuclei (if $a \ne b$), giving rise to three- and four-body correlations, respectively.
In our approach, the parameters $g_{\mu,\nu}^{a,b}$ of $J_{3/4}$ are
optimized by energy minimization, like the other variational parameters in the Jastrow and determinantal parts.

For the Jastrow atomic basis set $\chi^J$, we used linear combinations of Gaussian type orbitals (GTOs). The basis set depends on the atomic species.
For the carbon sites, we used the following contracted radial
basis functions: 
\begin{equation}
\begin{cases} 
\beta_1 e^{Z_1\cdot r^2}+\beta_2 e^{Z_2 \cdot r^2}& \text{for s orbitals}, \\
\beta_3 e^{Z_3\cdot r^2}+\beta_4 r e^{Z_4 \cdot r^2} & \text{for p orbitals}, \\
e^{Z_5\cdot r^2} & \text{for d orbitals},
\end{cases}
\end{equation}
where $\beta_i$ are the contraction coefficients and $Z_i$ the
Gaussian exponents, all variational parameters fully optimized during a
QMC energy minimization.
Analogously, the contracted Jastrow basis set for the hydrogen radial functions reads:
\begin{equation}
\begin{cases} 
\beta_1 e^{Z_1\cdot r^2}+\beta_2r e^{Z_2 \cdot r^2}& \text{for s orbitals}, \\
\beta_3 e^{Z_3\cdot r^2}+\beta_4 r e^{Z_4 \cdot r^2}& \text{for p orbitals}. 
\end{cases}
\end{equation}

The determinantal part $\Delta$ of $\Psi$ is critical
for this work, since its different forms will allow us to study the
physical consequences of electron correlation in the acenes. In
particular, it will tell us the nature of their ground state, as we
will illustrate in Sec.~\ref{acenes:results}. 
We are going to use three main 
forms for $\Delta$, namely the standard Slater form, yielding a
Jastrow Single Determinant (JSD) wave function, a fully resonant AGP
form, developed over the whole set of single particle
orbitals, leading to the JAGP wave function, and an intermediate case,
containing mainly HOMO-LUMO resonances, giving rise to a wave function
dubbed as ``Jastrow Double Determinant'' (JDD).
These three forms will be described in the next subsections.

With the conventional choice of $N^\uparrow>N^\downarrow$, the general
form of $\Delta$ is given by the antisymmetrization of  the matrix
$\phi_{ij}$, as follows:
\begin{equation}
\Delta(\textbf{r}_1^\uparrow,...,\textbf{r}^\uparrow_{N^\uparrow},\textbf{r}_1^\downarrow,...,\textbf{r}^\downarrow_{N^\downarrow})= \det(\phi_{ij}),
\label{acenes:eq:det}
\end{equation}
with
\begin{equation}
\phi_{ij}= 
\begin{cases} \phi (\textbf{r}_i^\uparrow,\textbf{r}_j^\downarrow) & \text{for j $\le$ $N^\downarrow$}, \\
\bar{\varphi}_j(\textbf{r}_i^\uparrow) & \text{for j $>$ $N^\downarrow$}.
\end{cases}
\label{acenes:eq:gem_with_spin}
\end{equation}
The $\phi$ function is the geminal, which pairs
opposite spin electrons in a spin-singlet state.
The determinant of $\phi$ in Eq.~\ref{acenes:eq:det} is called antisymmetrized geminal power (AGP)\cite{hurley1953,bratovz1965,bessis1967,goscinski1982}.
Spin polarization, if present, is given by the lone molecular orbitals
$\bar{\varphi_j}$ in Eq.~\ref{acenes:eq:gem_with_spin}, hosting unpaired electrons.

Both $\phi$ and $\bar{\varphi}_j$ are expansions over the atomic basis set $\chi$.
The basis set for $\Delta$ will be in general different
than the $\chi^J$ of the Jastrow factor. The geminal is a symmetric
quadratic form, and the $\bar{\varphi}_j$s are linear combinations of $\chi$, all
functions of spatial electron coordinates. They read
\begin{equation}
\phi(\textbf{r},\textbf{r}') = \sum\limits_{i,j}  \lambda_{i,j}
\chi_i(\textbf{r})\chi_j(\textbf{r}^\prime) 
\label{acenes:eq:atomic_expansion}
\end{equation}
for the geminal, and
\begin{equation}
\bar{\varphi}_\alpha(\textbf{r}) = \sum\limits_i  c^\alpha_i\chi_i(\textbf{r})
\label{acenes:eq:isolatedfun}
\end{equation}
for the unpaired orbitals. The $\lambda_{i,j}$ and $c^\alpha_i$ linear parameters
are optimized by energy minimization, with $i$ and $j$ running over the basis set elements $\chi_i$, 
centered on the nuclear coordinates $\textbf{q}_i$, analogously to $\chi^J_i$.
For $\chi$ we chose a contracted GTOs set, 
common to both the geminal and the unpaired 
orbitals. It is made of $(5s6p)/[3s2p]$ functions centered on the carbon sites and 
 $(5s)/[2s]$ functions centered on the hydrogen atoms. The QMC 
variational optimization scheme allows us to optimize not only the
linear coefficients but also the exponents 
of the Gaussian functions, considerably reducing the basis size needed
to reach the basis set convergence. As well known\cite{petruzielo2010}, its convergence is further speeded up by the simultaneous optimization of the Jastrow factor. 

The rank of the geminal function in Eq.~\ref{acenes:eq:gem_with_spin} sets the
level of static correlations included in $\Delta$. If its rank is equal
to $N_\downarrow$, one recovers the single Slater determinant. If the
rank is increased, static correlation is included. When it is the maximum allowed by the atomic basis set $\chi$, one will get the fully resonant JAGP form.
Owing to our local
basis set representation, this corresponds to the RVB ansatz,
with $\lambda_{i,j}$ defining the bond between
orbitals $\varphi_i$ and $\varphi_j$. 

The specific forms taken by $\Delta$ in this work are detailed below in an
increasing order of complexity.

\subsubsection{Jastrow Single Determinant (JSD)}

The geminal in Eq.~\ref{acenes:eq:gemdiag} can be written in the diagonal form
\begin{equation}
\phi(\textbf{r},\textbf{r}') = \sum\limits_{i}  \lambda_i \bar{\varphi}_i(\textbf{r})\bar{\varphi}_i(\textbf{r}'),
\label{acenes:eq:gemdiag}
\end{equation}
where the number of non-zero $\lambda_i$ values is equal to the
rank of the initial $\lambda_{i,j}$ matrix,
and the $\bar{\varphi}_i$ functions are molecular orbitals (MOs), linear combinations of $\chi$,
like the unpaired orbitals in Eq.~\ref{acenes:eq:gem_with_spin}.
The MOs generated by using the algorithm introduced in
Refs.~\onlinecite{michele_agp2} and \onlinecite{dupuy_vertical_2015}
are orthonormal, and multiplied by the coefficient $\lambda_i$, which
weights their contribution in the original AGP expansion.
The geminal function in Eq.~\ref{acenes:eq:gemdiag} provides
MOs that can be ordered according to $|\lambda_i|$. Thanks to the geminal algebra\cite{coleman1965,goscinski1982},
the eigenvectors $\bar{\varphi}_i$ of $\phi$ are the same as the ones of the one-body reduced density matrix (1-RDM) corresponding to the AGP $\Delta$. Therefore, the eigenvalues $\lambda_i$ are directly related to the natural orbital occupations, and
the orbitals $\bar{\varphi}_i$ that diagonalize the geminal in Eq.~\ref{acenes:eq:gemdiag} are the AGP natural orbitals (NOs).

By restricting the sum in Eq.~\ref{acenes:eq:gemdiag} to the first $N^\downarrow$ MOs with the largest $|\lambda_i|$
coefficients in the $\phi$ expansion, we obtain the following
geminal function
\begin{equation}
\phi^\textrm{SD}(\textbf{r},\textbf{r}') = \sum\limits_{i=1}^{N^{\downarrow}}  \lambda_i \bar{\varphi}_i(\textbf{r})\bar{\varphi}_i(\textbf{r}'),
\label{acenes:eq:gemdiag2}
\end{equation}
whose determinant leads in Eq.~\ref{acenes:eq:det} to the standard single
Slater Determinant for both singlet and spin polarized wave
functions. For singlet wave functions, this corresponds to keeping all
MOs until the HOMO (rank-$N^\downarrow$ geminal). The Jastrow
correlated wave
function having $\Delta$ generated in this way is our best
representative of a closed-shell singlet (CSS) form for the acenes ground state.

On the other hand, the SD triplet state can be generated by the
same Eq.~\ref{acenes:eq:gemdiag2}, by including a summation of
MOs until the HOMO-1, while HOMO and LUMO will be lone orbitals occupied by unpaired
electrons (Eq.~\ref{acenes:eq:gem_with_spin}).

\begin{figure}[h]
\centering 
      \includegraphics[width=1.0\columnwidth]{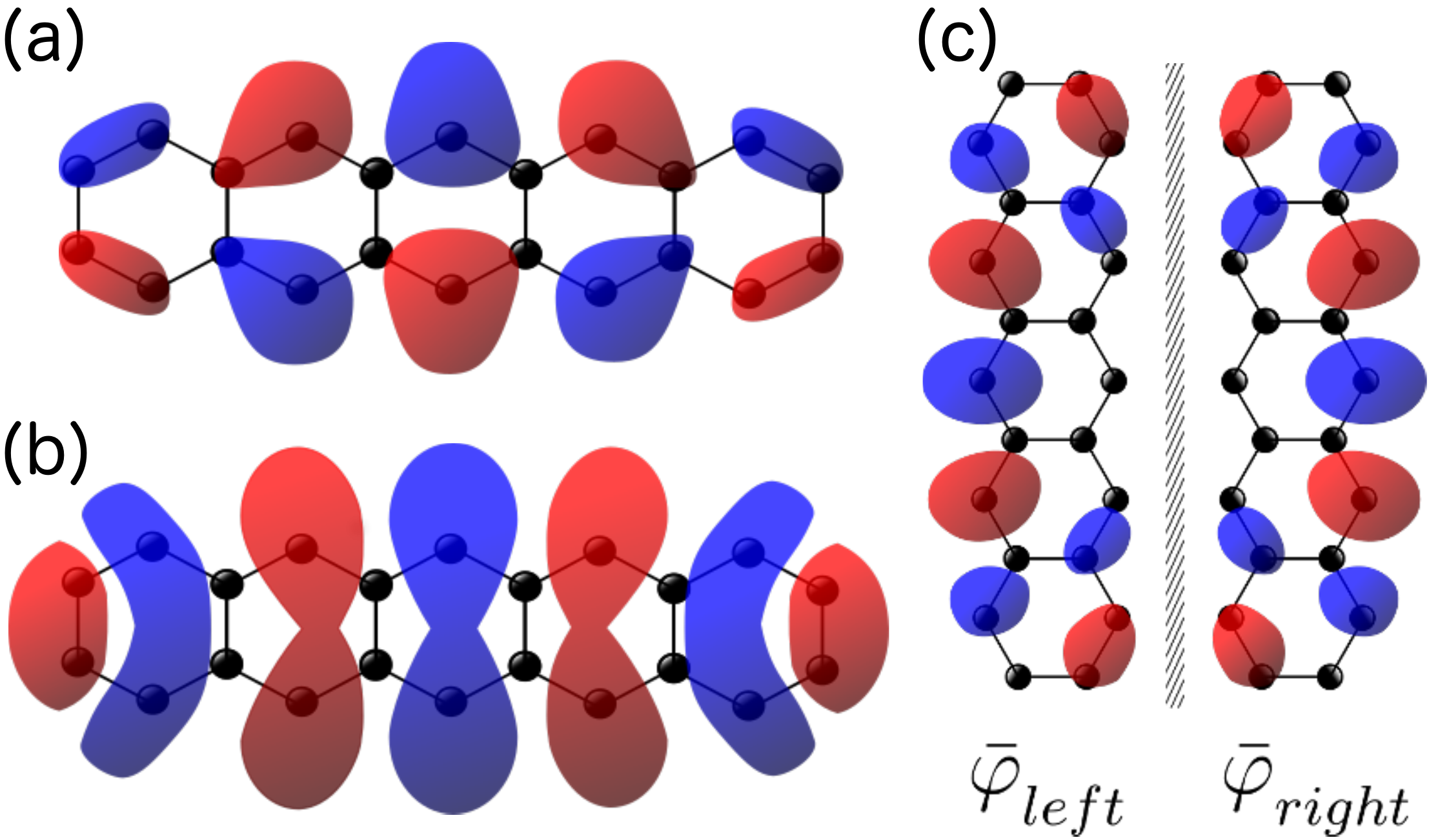} 
\caption{Panel (a): HOMO isosurface of the pentacene molecule as obtained by LDA calculations in a Gaussian basis set. Panel (b): as in panel (a), but for the LUMO. Panel (c): edge orbitals in pentacene obtained as symmetric and antisymmetric linear combinations of HOMO and LUMO.\label{acenes:figure:orbital_collage}
}
\end{figure}

We plot the HOMO and LUMO orbitals for the pentacene molecule in
Fig.~\ref{acenes:figure:orbital_collage}, panels (a) and (b). They show a $b_{3g}$
and $b_{1u}$ symmetry, respectively, in agreement with previous works\cite{hashimoto_theoretical_1996}.
HOMOs and LUMOs exhibit the same symmetry pattern ($b_{3g}$
and $b_{1u}$ for odd-ring chains, $a_u$ and $b_{2g}$ for even-ring acenes) in the whole
set of investigated molecules.  The HOMO-LUMO gap gets smaller as the
acene length increases. This gap reduction is
closely related to the possibility of a singlet-triplet instability,
or to the occurrence of an open-shell singlet for long-enough chains, as we mentioned in the Introduction.

\subsubsection{Jastrow Double Determinant (JDD)}

To address the question about whether an open-shell singlet (OSS) becomes the
ground state of long-enough acene chains, with strong electron correlation
between the long edges, we need to define an appropriate variational
ansatz for the OSS. As apparent from Bendikov's paper\cite{bendikov2004oligoacenes},
the open-shell orbitals are localized on the molecular long edges. We
can therefore dub them without ambiguity as $\bar\varphi_{left}$ or
$\bar\varphi_{right}$, depending on their left or right edge localization.
Thus, the open-shell singlet wave function, that does not break the spin
symmetry, must have the following form:
\begin{eqnarray}
\Delta_\textrm{OSS} (\textbf{r}^\uparrow,
\textbf{r}^\downarrow) & = & \frac{1}{\sqrt{2}} \left( \left|
    \bar\varphi_{left}(\textbf{r}^\uparrow)
    \bar\varphi_{right}(\textbf{r}^\downarrow) \right|
                             \right. \nonumber \\ 
& + & \left. \left|
    \bar\varphi_{right}(\textbf{r}^\uparrow)
    \bar\varphi_{left}(\textbf{r}^\downarrow) \right| \right), 
\label{acenes:OSS_2det}
\end{eqnarray}
where for the sake of clarity we omitted the other lower energy states
($\{\bar\varphi_1, \ldots,  \bar\varphi_\textrm{HOMO-1} \}$) doubly occupied. As usual, the $|\cdots|$ indicates a Slater determinant.

In our paper, we are going to use a more general wave function than the one in
Eq.~\ref{acenes:OSS_2det}, which will transform into the exact OSS form in a
particular limit. Our ansatz is based on a geminal form, which can
easily interpolate between the CSS and OSS states. 
As explained in Sec.~\ref{acenes:intro}, the physics behind the open-shell
singlet ground state stabilization is related to the small or
vanishing HOMO-LUMO gap. We expect, therefore, that the leading
contributions to static correlation will come from the HOMO and LUMO
states, which must both be included in the variational ansatz.
This is obtained by summing the geminal expansion up to the LUMO orbital (rank-($N^\downarrow+1$) geminal): 
\begin{eqnarray}
\phi^\textrm{DD}(\textbf{r},\textbf{r}') & = &
\sum\limits_{i=1}^{N^\downarrow-1} \lambda_i
\bar{\varphi}_i(\textbf{r})\bar{\varphi}_i(\textbf{r}') \nonumber \\ 
&+ & \lambda_\textrm{HOMO}\bar{\varphi}_\textrm{HOMO}(\textbf{r})\bar{\varphi}_\textrm{HOMO}(\textbf{r}')
  \nonumber \\
&+ & \lambda_\textrm{LUMO}\bar{\varphi}_\textrm{LUMO}(\textbf{r})\bar{\varphi}_\textrm{LUMO}(\textbf{r}'),  
\label{acenes:eq:gemdiag_JDD}
\end{eqnarray}
The antisymmetrization of $\phi^\textrm{DD}$ in Eq.\ref{acenes:eq:det}
generates $N^\downarrow+1$ Slater determinants, which can be grouped in three types: 
\begin{widetext}
\begin{equation}
\begin{cases}
|\bar\varphi_\textrm{HOMO}^{\uparrow}\bar\varphi_\textrm{HOMO}^{\downarrow}\prod\limits_{i=1}^{N^{\downarrow}-1} \bar\varphi_i^{\uparrow}\bar\varphi_i^{\downarrow}|\textrm{~~~(type I, 1 occurrence)}, \\
|\bar\varphi_\textrm{LUMO}^{\uparrow}\bar\varphi_\textrm{LUMO}^{\downarrow}\prod\limits_{i=1}^{N^{\downarrow}-1} \bar\varphi_i^{\uparrow}\bar\varphi_i^{\downarrow}|\textrm{~~~(type II, 1 occurrence)}, \\
|\bar\varphi_\textrm{HOMO}^{\uparrow}\bar\varphi_\textrm{HOMO}^{\downarrow}\bar\varphi_\textrm{LUMO}^{\uparrow}\bar\varphi_\textrm{LUMO}^{\downarrow}\prod\limits_{\substack{i=1 \\ i \neq k}}^{N^{\downarrow}-1} \bar\varphi_i^{\uparrow}\bar\varphi_i^{\downarrow}|\textrm{~~~(type III, $N^\downarrow-1$ occurrences)}, \\
\end{cases}
\end{equation}
\end{widetext}
where in the last case, the $k$-th orbital is excluded from the
product of the remaining MOs. The first type is the reference
state occupation, the second one is a state obtained by a HOMO-LUMO 
pair excitation, the third
one is made of pair excitations to the LUMO from lower-energy orbitals.
All those configurations are weighted
by the product of their respective geminal expansion coefficients:
\begin{equation}
\begin{cases}
\lambda_\textrm{HOMO}\prod\limits_{i=1}^{N^{\downarrow}-1} \lambda_i\textrm{~~~for type I}, \\
\lambda_\textrm{LUMO}\prod\limits_{i=1}^{N^{\downarrow}-1} \lambda_i\textrm{~~~for type II}, \\
\lambda_\textrm{HOMO}\cdot\lambda_\textrm{LUMO}\prod\limits_{\substack{i=1 \\ i \neq k}}^{N^{\downarrow}-1} \lambda_i\textrm{~~~for type III}. \\
\end{cases}
\end{equation}

In the limit of all $\lambda_i$ values equal to $1$ for
$i=\{1,\ldots,\textrm{HOMO-1}\}$, and the two other
coefficients  $|\lambda_\textrm{HOMO}|,|\lambda_\textrm{LUMO}| \ll 1$, then the
prefactors weighting the type III become negligible relative
to the other two, and $\phi^\textrm{DD}$ gives rise to a pure HOMO-LUMO resonance.
In this situation, the OSS state emerges when $\lambda_\textrm{HOMO}=- \lambda_\textrm{LUMO} = \epsilon $ in the limit of small $\epsilon $. 
Indeed, one can rewrite
\begin{equation}
\begin{split}
&\epsilon \cdot
\bar\varphi_\textrm{HOMO}^{\uparrow}\bar\varphi_\textrm{HOMO}^{\downarrow} -
\epsilon \cdot
\bar\varphi_\textrm{LUMO}^{\uparrow}\bar\varphi_\textrm{LUMO}^{\downarrow} =\\
&\epsilon\cdot\bar\varphi_{left}^{\uparrow}\bar\varphi_{right}^{\downarrow} + \epsilon\cdot\bar\varphi_{right}^{\uparrow}\bar\varphi_{left}^{\downarrow},
\end{split}
\end{equation}
once one defines the left and right orbitals as linear combinations of
HOMO and LUMO, as follows:
\begin{eqnarray}
\bar\varphi_{left} & = &  \frac{1}{\sqrt{2}} \left( \bar\varphi_\textrm{HOMO} +
                         \bar\varphi_\textrm{LUMO} \right), \nonumber \\
\bar\varphi_{right} & = & \frac{1}{\sqrt{2}} \left( \bar\varphi_\textrm{HOMO} - \bar\varphi_\textrm{LUMO} \right). 
\end{eqnarray}

Owing to the spatial symmetry of the HOMO and LUMO states, their symmetric
and antisymmetric combinations lead to states localized on the edge of
the molecule. It turns out
that within this representation,  $\bar\varphi_{left}$ and
$\bar\varphi_{right}$ are mirror images of each other, through the
reflection with respect to the plane containing the long axis and
perpendicular to the molecule (see Fig.~\ref{acenes:figure:orbital_collage}(c)). In this limit, our ansatz reduces to
the one in Eq.~\ref{acenes:OSS_2det}, containing only two Slater
determinants.

On the other hand, for $\{\lambda_i=1\}_{i=1,\ldots,\textrm{HOMO}}$, and
$\lambda_\textrm{LUMO}=0$, one trivially recovers the closed-shell single
Slater determinant. Thus, both CSS and OSS are included in the JDD ansatz, which can interpolate between the two.

To optimize the previously described JSD and JDD wave 
functions, the $\lambda_{i,j}$ coefficients 
are constrained at a fixed rank of the geminal matrix, and diagonalized at each energy minimization step to get new MOs.
This scheme has already been employed 
in the QMC calculation of the low-lying energy spectrum of the anthracene molecule\cite{dupuy_vertical_2015}, and in the Ref.~\onlinecite{zen_static_2014} for the energy optimization of diradical molecules.

\subsubsection{Fully resonant Jastrow Antisymmetric Geminal Power (JAGP)}

According to Eq.~\ref{acenes:eq:atomic_expansion}, the fully resonant JAGP wave function is obtained when the geminal is 
expanded in the full Hilbert space spanned by the GTO basis set $\chi$:
\begin{equation}
\phi^\textrm{AGP}(\textbf{r},\textbf{r}^\prime) = \sum\limits_{i,j}  \lambda_{i,j}
\chi_i(\textbf{r})\chi_j(\textbf{r}^\prime),
\label{acenes:eq:geminal_full}
\end{equation}
with $\lambda_{i,j}$ expansion coefficients. In other words,
the rank of the $\phi^\textrm{AGP}$ is not
limited (its upper bound is set just by the basis set
extension). Therefore, the JAGP has the largest
variational freedom among our three classes of wave functions. This
framework is the closest to the valence bond picture, where each
chemical bond is described by a set of $\lambda_{i,j}$, involving the
$\chi_i$ and $\chi_j$ atomic orbitals. In this picture,
the spatial dependence of correlations in the system is
transparent, because it is directly related to the variation of
$\lambda_{i,j}$ as a function of internuclear distance, thanks to the
localized GTO basis set. One can always convert the JAGP
into the MO picture, by diagonalizing the
geminal $\phi^\textrm{AGP}$. In this case, one
will find a full set of NOs, including all virtual
states. Studying the behavior of their occupations in the JAGP
quantifies the role of static
correlations in the system.

\subsection{Technical details}
\label{acenes:technical}

We ran preliminary LDA calculations in the same GTO basis set as the one of QMC calculations (i.e. $(5s6p)/[3s2p]$ for carbon, and $(5s)/[2s]$ for hydrogen), to generate the initial molecular orbitals. Both LDA and QMC calculations have been done by running the TurboRVB suite of codes\cite{sorella_rvb}. We optimized the QMC wave function by using the ``linear method'', which includes also partial information of the Hessian\cite{SR4}, in its iterative conjugate-gradient version\cite{michele_benzene}. The projective LRDMC simulations\cite{casula_diffusion_2005} have a lattice space of 0.25 $a_0$, randomly oriented\cite{filippi_lrdmc}, which gives converged results in the energy differences. The geometries have been optimized at the B3LYP functional by using the VASP code\cite{kresse_efficient_1996}, with plane waves and PAW pseudopotentials. The largest supercell size in VASP calculations is 60\AA $\times$30\AA$\times$10\AA, used for nonacene to minimize the periodic image bias. Restricted R-B3LYP calculations have been carried out for both the singlet and the triplet state. For the anthracene, whose closed-shell character is the largest, we verified that the R-B3LYP geometry is in good agreement with the QMC one, previously published in Ref.~\onlinecite{dupuy_vertical_2015}. For the nonacene, which is affected by strong static correlations, the geometries have further been optimized at the VMC level\cite{barborini2012structural}.

\section{Results}
\label{acenes:results}

\subsection{Singlet-Triplet gap}
\label{acenes:singlet-triplet_gap}

\begin{figure}[h]
\centering 
\includegraphics[width=1.0\columnwidth]{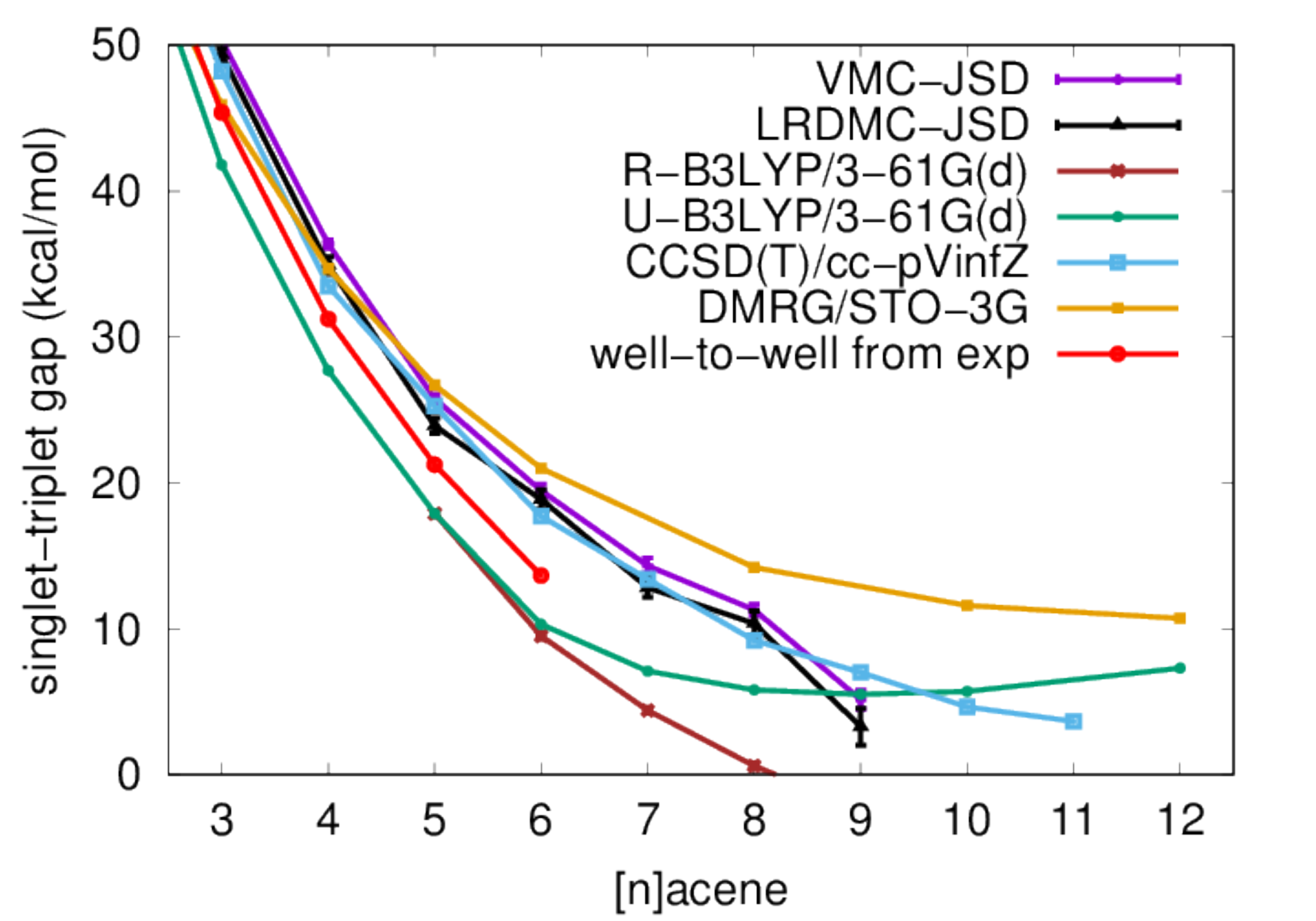} \\
\caption{Adiabatic (``well-to-well'') singlet-triplet gap as a function of the number of rings in the molecule, for different theories: variational and diffusion Monte Carlo with JSD wave function (VMC-JSD and LRDMC-JSD, this work), restricted and unrestricted B3LYP (R-B3LYP and U-B3LYP, Ref.~\onlinecite{bendikov2004oligoacenes}), CCSD(T) with focal point analysis\cite{hajgato2009benchmark,hajgato_focal_2011}, DMRG/CASCI in STO-3G minimal basis set (DMRG, Ref.~\onlinecite{chan_dmrg}), the experimental gap where the zero point energy (ZPE) contribution has been subtracted (ZPE and experimental values from Ref.~\onlinecite{hajgato_focal_2011}). 
\label{acenes:figure:singlet_triplet_gap}
}
\end{figure}

We start by studying the singlet-triplet gap as a function of the acene length. It is a significant quantity to take into account, because it is directly related to the HOMO-LUMO gap, and to its closure in the large acene limit. Indeed, the singlet-triplet gap involves a HOMO-LUMO excitation, accompanied by a spin flip. The HOMO-LUMO energy levels become degenerate in the infinite length limit. Therefore, the system is prone to a number of possible symmetry breakings, to gain energy by lowering its symmetry\cite{bettinger2010}. One of them is the triplet instability, predicted by Houk \emph{et al.}\cite{houk2001}, for a number of fused rings larger than 8. By carrying out B3LYP/6-31G* calculations, they found that the triplet state becomes more favorable in energy than the singlet for chains longer than octacene, with a remarkable bond equalization, namely the absence of bond length alternation. Experimentally, Angliker and coworkers\cite{angliker_electronic_1982} arrived to a similar conclusion already in 1982. They extrapolated excitation energies available at that time until the hexacene, and predicted that the nonacene would be the first polyacene showing a triplet spin state. The impossibility of synthesizing acenes larger than 6 rings, due to their extreme reactivity, prevented a direct test of these predictions on electronic spectra of longer molecules. The possible occurrence of spin instabilities has been claimed since the appearance of Hartree-Fock (HF) calculations\cite{baldo1983semiconductor}, first done on parametrized Pariser-Parr-Pople (PPP) -  aka extended Hubbard -  Hamiltonians, and later on realistic ones\cite{dehareng2000}. An obvious question is whether spin instabilities survive when higher-level theories are employed. HF theory can provide however some deep physical insight, suggesting the need of a bi- or multi-configurational reference. Based on these considerations, Bendikov \emph{et al.}\cite{bendikov2004oligoacenes} explored the existence of an antiferromagnetic solution, and its stability with respect to the high-spin triplet state. Performing spin unrestricted B3LYP/6-31G* calculations, they found that an open-shell singlet (OSS) solution, with spin-polarized orbitals lying on the long zig-zag edge of the polyacenes, are always more stable than the corresponding triplet states. Moreover, at the B3LYP level, the OSS energy is even lower than the closed-shell singlet (CSS) - i.e. the spin-restricted solution -, for acenes larger than pentacene. This is reflected in the behavior of the singlet-triplet gap, reported in Fig.~\ref{acenes:figure:singlet_triplet_gap}, by the raise of the unrestricted B3LYP curve for larger sizes. Therefore, according to Ref.~\onlinecite{bendikov2004oligoacenes}, the triplet spin configuration is \emph{not} the true ground state of higher acenes.

These results were confirmed by an impressive amount of subsequent works. The authors of Ref.~\onlinecite{jiang2008electronic} reached the same conclusions by using a different functional (the PBE generalized gradient approximation) and a different basis set (plane waves), always allowing for unrestricted (i.e. spin-polarized) solutions. The drawback of this type of solutions is their strong spin contamination, which increases with system size. The resulting wave function is far from being an eigenstate of $S^2$. Efforts to overcome this issue have been done within the DFT framework, in the fractional-spin (FS)\cite{ess2010} and the recently developed thermally-assisted-occupation (TAO)\cite{chai2012,chai2014,wu2016} variants. When the spin symmetry is restored in FS-DFT, the triplet state becomes lower in energy again, already for the octacene\cite{ess2010}. In contrast, TAO-DFT yields the singlet as the lowest energy state. The absence of spin contamination can be checked \emph{a posteriori} in TAO-DFT, by comparing the unrestricted and restricted energies, which turn out to be identical within numerical precision\cite{wu2016}. From the experimental side, new significant advances have been achieved in last years in the synthesis of larger acenes\cite{dorel2017}. Thanks to the development of new synthetic methods\cite{zade_heptacene_2010} and cryogenic matrix-isolation techniques\cite{mondal_synthesis_2009}, acenes as long as nonacene have been generated\cite{tonshoff2010photogeneration,zuzak2017}, even in crystalline forms\cite{purushothaman2011}. Thus, their structures and optical spectra are now accessible. By studying the evolution of the spectral patterns with the molecular length, T{\"o}nshoff and Bettinger provided a strong experimental support for the spin-singlet nature of the ground state up to 9 rings\cite{tonshoff2010photogeneration}. Theoretical calculations of optical absorption, performed at the multireference singles-doubles configuration interaction (MRSDCI) level, affordable on semi-empirical PPP Hamiltonian, gave a further strength to the experimental conclusions\cite{chakraborty2013}.

The unrestricted solutions point towards a multiconfigurational character of the true ground state, affected by strong static correlations. Complete active space (CAS) schemes can naturally deal with multiconfiguration wave functions. Unfortunately, the application of CAS approaches to the oligoacene is hard, due to the exponentially large active space one should include in the calculations for longer and longer molecules. However, thanks to the unidimensional nature of these systems, the density matrix renormalization group (DMRG) can be very effective to accelerate the CAS evaluation. DMRG/CASCI with an active space including all $\pi$ electrons and with a minimal STO-3G basis set has been carried out in Refs.~\onlinecite{chan_dmrg,mizukami2012}. These multi-reference calculations clearly gave the singlet as the true ground state, without breaking the spin symmetry, but with a singlet-triplet gap that is overestimated compared to the experiment (see Fig.~\ref{acenes:figure:singlet_triplet_gap}). 

Not only the static correlations, but also the dynamic ones could be important to stabilize the singlet with respect to the triplet for large sizes. Indeed, a subtle balance between dynamic and static correlations has been put forward in Ref.~\onlinecite{ibeji2015}, as revealed by spin-flip methods applied to correlated reference states. In fact, single-reference coupled cluster (CC) calculations, extrapolated in both basis set and theory according to a focal point analysis, provide more accurate results when compared with experiments\cite{hajgato2009benchmark}, as shown in Fig.~\ref{acenes:figure:singlet_triplet_gap}. Also in this case there is no triplet instability, as explicitly computed at the CCSD(T)/cc-pV$\infty$Z level of theory up to undecacene\cite{hajgato_focal_2011}. The extrapolation of those data gives a vanishing singlet-triplet gap in the infinite size limit, within an uncertainty of 1.5 kcal/mol. This would confirm the HOMO-LUMO gap closure for the infinite chain by single-reference correlated calculations.

\begin{table}[ht]
\begin{tabular}{ r | r | r}
 $n$-acene  & JSD-VMC gap (kcal/mol)  & JSD-LRDMC gap (kcal/mol) \\
\hline
3 & -50.33 (0.45)  &    -49.33 (0.61) \\ 
4 & -36.37 (0.43)   &    -34.93 (0.83)  \\
5 & -25.78 (0.40)   & -23.90 (0.75)  \\
6 & -19.46 (0.61)  &  -18.86 (0.95) \\
7 & -14.31 (0.74)   &   -12.82 (0.96)  \\
8 & -11.29 (0.57)  &   -10.35 (1.22) \\
9  & -4.19 (0.55)   &  -3.28 (1.79) \\
\end{tabular}
\caption{Adiabatic singlet-triplet gap computed at the VMC and LRDMC levels for JSD wave functions. The zero point energy is not included in the ``well-to-well'' estimates.
\label{acenes:table:singlet_triplet_gap}
}
\end{table}

Our QMC calculations based on the JSD wave function yield a singlet-triplet gap in good agreement with the CCSD(T) values. Our results are reported in Tab.~\ref{acenes:table:singlet_triplet_gap} and Fig.~\ref{acenes:figure:singlet_triplet_gap} for both variational (VMC) and lattice regularized diffusion Monte Carlo (LRDMC) simulations. They are based on a single-reference wave function, because the Jastrow factor correlates a single Slater determinant. Therefore, it is not surprising that they agree with CCSD(T), as they describe the same physics. However, a direct comparison between the two theories is worth it. It reveals that our Jastrow factor is able to successfully capture a large amount of dynamic correlation, yielding an accuracy comparable with CCSD(T)/cc-pV$\infty$Z already at the VMC level. In Fig.~\ref{acenes:figure:singlet_triplet_gap}, one can note that the QMC gaps fluctuate more for larger  $n$-acenes. This could be related to an ``even-odd'' alternation effect more pronounced in QMC than in CCSD(T), combined with larger statistical fluctuations in the total energy, due to the increasing size of the system.

\begin{figure}[h]
\centering 
\includegraphics[width=1.0\columnwidth]{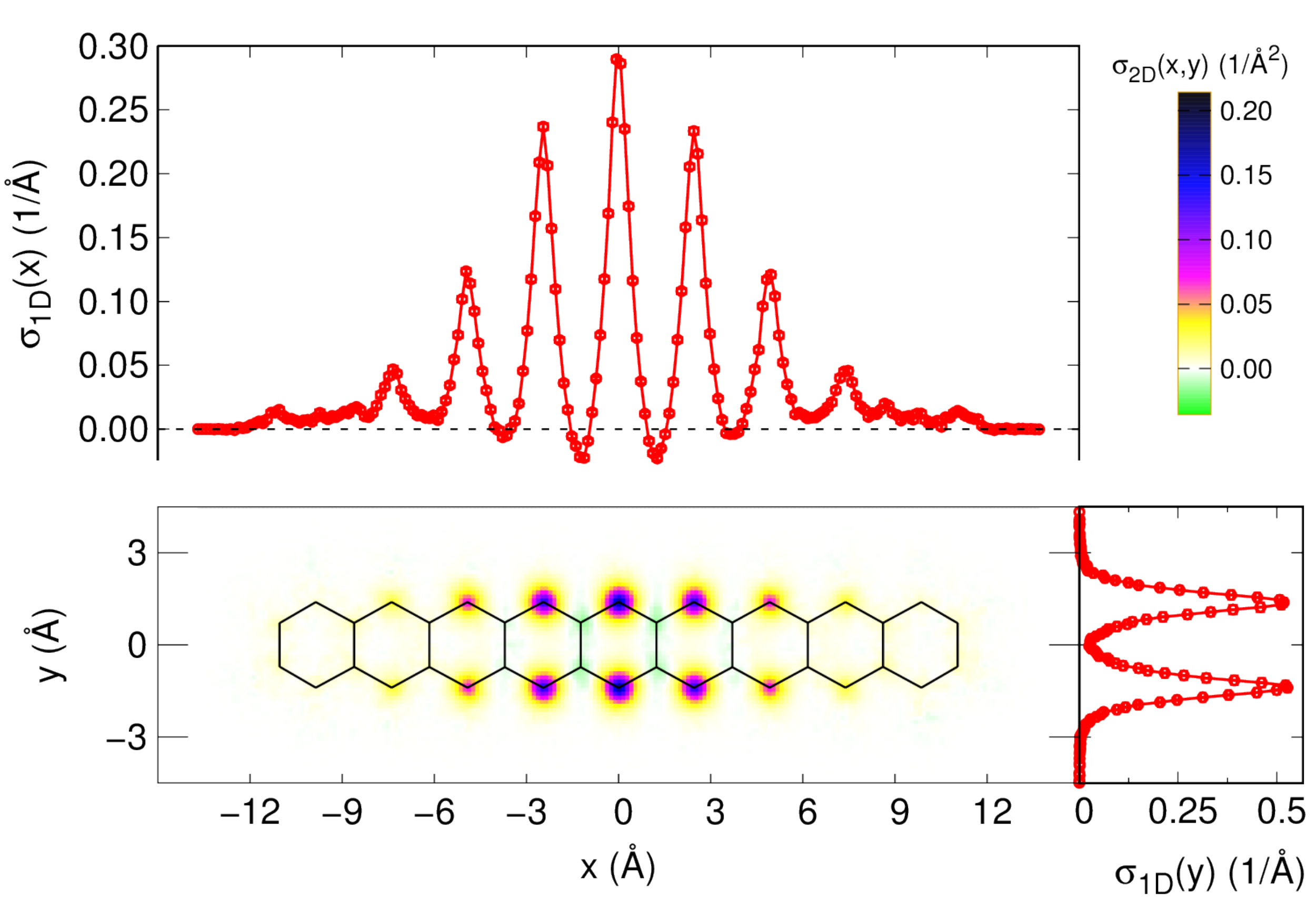} \\
\caption{\label{acenes:figure:triplet_spin_density} Spin density $\sigma$ from VMC-JSD calculations of nonacene in its triplet state. 
Bottom left panel: contour plot of the spin density (the corresponding color code is in the top right box). Top left panel: spin density projection on the long axis of the molecule. Bottom right panel: spin density projection on the short molecular axis. 
}
\end{figure}

Despite not being the true ground state of the system, the triplet wave function is very useful to analyze, in order to probe the localization of the unpaired electrons along the two edges of the oligoacenes. Indeed, at the leading order, the main difference with respect to the OSS is the spin occupation of the edge-localized \emph{right} and \emph{left} orbitals (Fig.~\ref{acenes:figure:orbital_collage}(c)), ordered ferromagnetically in the triplet, antiferro in the OSS. Therefore, the orbital localization can be very easily detected in the triplet state, as it is directly given by the spin density, which is plotted in Fig.~\ref{acenes:figure:triplet_spin_density} for the nonacene. We have chosen the longest molecule in our study, because it is supposed to be the most critical one to show strong edge localization.

\begin{figure}[h]
\centering 
\includegraphics[width=1.0\columnwidth]{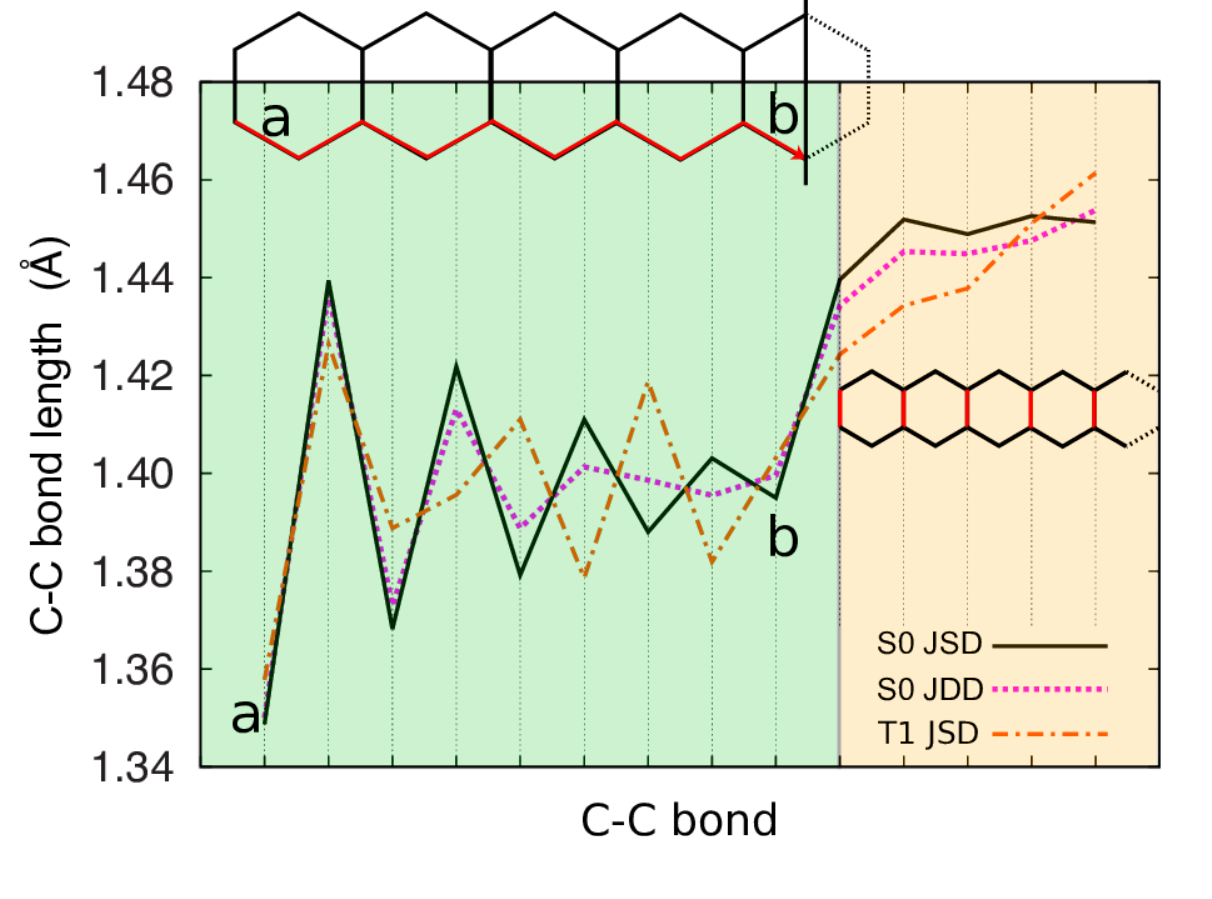} \\
\caption{C-C bond lengths of nonacene in the JSD and JDD singlets and JSD triplet states, after geometry relaxation performed at the VMC level. Left panel: bond lengths on the zig-zag frontier along the path from $a$ the outermost bond) to $b$ (the closest to the vertical symmetry axis). Right panel: rung bond lengths plotted from the outermost vertical bond to the most central one. The position of the bonds is visualized by red segments, vertically aligned with the corresponding length value points. 
\label{acenes:figure:triplet_vs_singlet_geo}
}
\end{figure}

As one can evince from Fig.~\ref{acenes:figure:triplet_spin_density}, the two like-spin electrons are localized in both longitudinal (upper panel) and transverse (bottom right panel) directions. The singlet-triplet gap is related to the overlap between \emph{right} and \emph{left} orbitals. In the transverse direction, the overlap between the two edge states is very small. The electrons are mostly localized in the outer carbon sites of the zig-zag edge, and in particular in the middle of the molecule. This electron arrangement has a strong impact on the equilibrium
geometry. To study the interplay between electronic configuration and molecular structure, we relaxed the geometry at the VMC level.
The result is shown in Fig.~\ref{acenes:figure:triplet_vs_singlet_geo}, where we compare the CSS relaxed geometry
with the one obtained by relaxing the triplet. The difference between the two is striking. The periodicity of the bond length alternation (BLA) in the center of the molecule is reversed, due to the substantial weakening of the resonating bonds. This can be explained within Clar's theory, by the breaking of Kekul\'e resonances in the central rings due to the pinning of isolated electrons at the outer edges, and the introduction in Clar's diagrams of long bonds made of weakly non-neighbor-paired electrons\cite{bhattacharya_clar_2014}. The phenomenon is manifested also by the elongation of the central rung bond. At the terminations of the chain, the in-phase periodicity of the BLA between the singlet and the triplet states is instead restored, owing to strong boundary effects. The charge, rather than spin, localization is the main responsible for the out-of-phase behavior of BLA in the central rings. However, in the triplet state the localization effects are certainly enhanced, as a consequence of the Pauli repulsion between like-spin particles, which pulls the edge electrons further apart. From our analysis, it turns out that not only the electron density profile, hardly accessible by experiments, but also the geometry is a probe of electron localization in the acenes.

In this Section we have seen that explicitly correlated theories do not need to break the spin symmetry to yield singlet energies lower than the triplet in higher acenes, in contrast to ``regular'' DFT-based calculations. Dynamic correlations, as included in CC and JSD-QMC, are enough to stabilize a spin-singlet ground state. In the next subsection, we will find out \emph{which} spin singlet is the most likely ground state of the oligoacenes, according to our QMC calculations.

\subsection{Open- and closed-shell singlets}
\label{acenes:open_closed-shells}

As we have mentioned in Sec.~\ref{acenes:singlet-triplet_gap}, the instability of the B3LYP and PBE functionals towards a symmetry-broken spin-unrestricted solution is the signature of a multi-reference character of the oligoacenes ground state. This has been confirmed by high-level methods\cite{chan_dmrg,yang2016}, applied to longer acenes to verify Bendikov's predictions on the OSS ground state. The possibility for the wave function to possess a multiconfigurational character is essential, if one wants to describe an OSS situation. In our QMC framework, we need to go beyond the
single-reference description of the JSD wave function, by adding static correlations. To do so, we employed two variational wave functions: the JDD, and the JAGP/RVB forms. The former explicitly includes HOMO-LUMO excitations, while the latter is the wave function representation of the RVB ansatz, proposed by
Pauling and Wheland as the ideal candidate to describe aromatic compounds\cite{pauling1933nature}. On the other hand, the single-reference JSD wave function represents a perfect CSS state, with only dynamic correlations.

Using the JDD and RVB variational forms allows us to study the evolution of the multi-reference character as a function of the molecular size. Indeed, the JDD wave function is able to describe a diradical situation, with two electrons perfectly localized at the acene edges, one on each side (OSS involving two electrons). As pointed out by many authors\cite{yang2016}, this can be mapped into the paradigmatic case of the H$_2$ molecule in the dissociation limit. As it is well known, the JDD ansatz can describe very accurately this situation\cite{Casula2004,michele_agp2,zen_static_2014}. It is also known that by restricting the variational freedom to the leading electron excitations to the LUMO state, there is a tendency of enhancing the diradical character. Instead of being a drawback, we turn it into our advantage, for we can learn much more about the true nature of the ground state, by comparing the JDD and RVB properties on an equal footing. The perfect OSS configuration with two unpaired electrons (perfect diradical) is obtained within JDD when $\lambda_\textrm{HOMO}=-\lambda_\textrm{LUMO}$, as explained in Sec.~\ref{acenes:wfs}.

On the other hand, the JAGP/RVB wave function includes all possible valence bond singlets, arranged in the $sp^2$ network. It is an analytically compact representation of the linear combination of all Clar's diagrams, whose number grows exponentially with system size. 
The AGP part is written in a localized atom-centered basis set, and is developed on the full network, by linking in a pairwise manner all sites via the $\lambda_{i,j}$ couplings (see Sec.~\ref{acenes:wfs}). The Jastrow factor applied on the AGP is supposed to select only those valence bond configurations that do not have double $p_z$ occupancies on the carbon atoms. Therefore, in the JAGP the Jastrow factor not only deals with dynamic correlations, such as the $\sigma$-$\pi$, dispersive (long-range) and Coulomb hole (short-range) electron-electron correlations, but also acts as a valence bond configurations filter, by keeping the most relevant ones that enter the Clar's expansion. In this sense, it acts as a Gutzwiller projector. The strength of its projection is set by the wave function optimization based on the VMC energy minimization. 
Therefore, the JAGP wave function representation overcomes one of the major issues common in CAS approaches and explicit valence bond theories\cite{wu2011}, when applied to the PAHs. Indeed, as the system size increases, one has to include a larger and larger number of states in the active space, in order to keep an adequate accuracy. This number becomes soon intractable, unless one resorts to approximations\cite{pelzer2011,fosso2016,schriber2016,battaglia2017} or uses DMRG methods by exploiting the low dimensionality of the system\cite{chan_dmrg,mizukami2012}. Instead, in the JAGP/RVB ansatz, an exponentially large number of resonating valence bond configurations is kept in a polynomial $L^3$ cost, with a set of variational parameters which grows only as $L^2$ with the system size $L$.

\begin{figure}[h]
\centering 
\includegraphics[width=1.0\columnwidth]{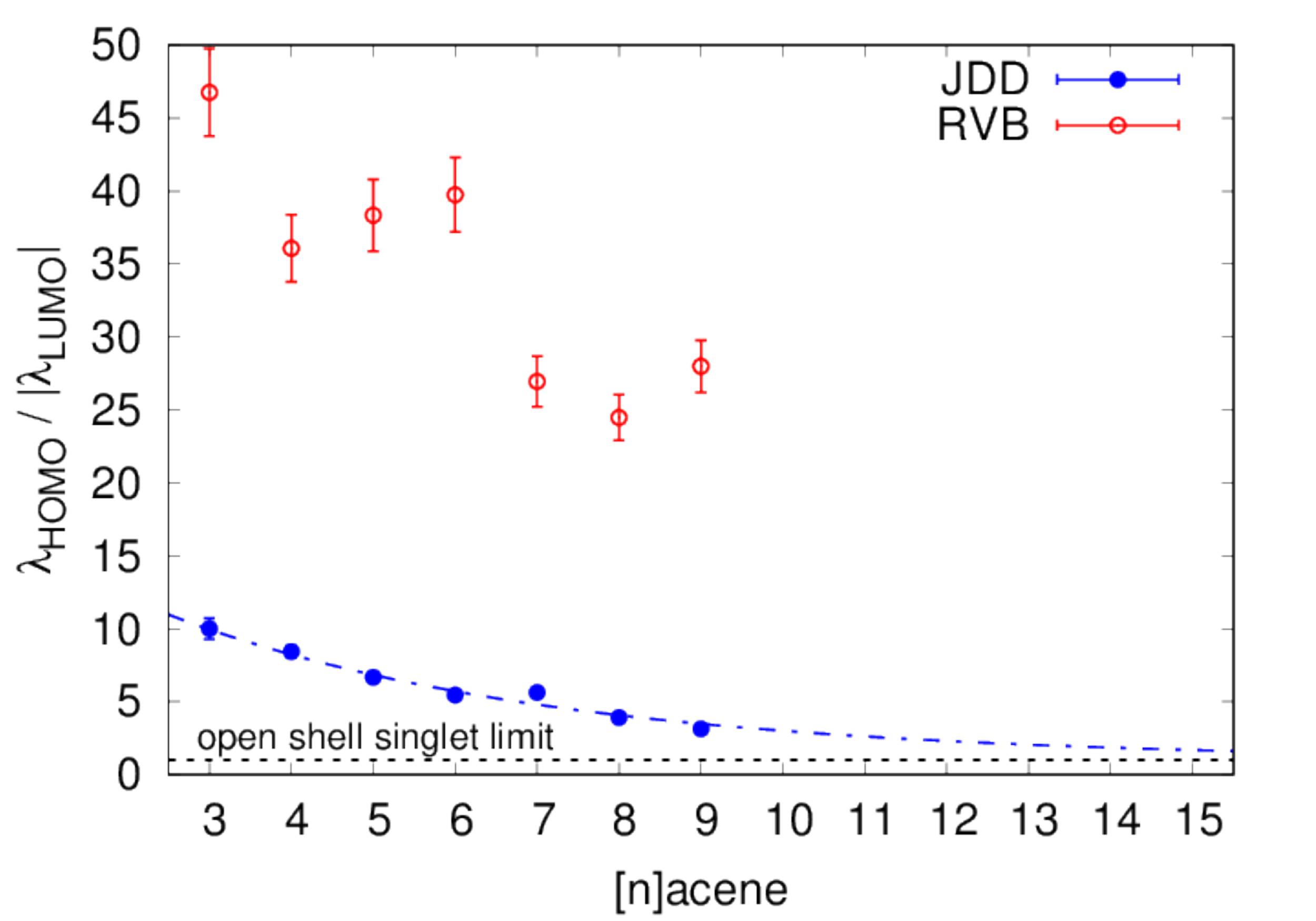} \\
\caption{\label{acenes:figure:lambda_H2L}
$\lambda_\textrm{HOMO}/|\lambda_\textrm{LUMO}|$ ratio, where $\lambda_\textrm{HOMO}$ and $\lambda_\textrm{LUMO}$ are the HOMO and LUMO weights, respectively. This is plotted as a function of the total number of rings in the corresponding $n$-acene molecule. Both JDD and RVB values are reported. The open-shell-singlet limit (dashed line) corresponds to $\lambda_\textrm{HOMO} = -\lambda_\textrm{LUMO}$. The JDD values are interpolated by the function $f(n)=1+\alpha \exp(-\beta n)$, with $\alpha=17$ and $\beta=0.214$ best fitting parameters (dotted-dashed line). 
}
\end{figure}

In Fig.~\ref{acenes:figure:lambda_H2L}, we plot the $\lambda_\textrm{HOMO}/|\lambda_\textrm{LUMO}|$ ratio, using the optimal parameters after the JDD and RVB energy minimizations. 
In the JDD wave function, the $\lambda_\textrm{HOMO}$ and $\lambda_\textrm{LUMO}$ - the HOMO and LUMO weights - are direct variational parameters. In the JAGP/RVB ansatz, the highest occupied natural orbital (HONO) and lowest unoccupied natural orbital (LUNO), together with their respective weights, $\lambda_\textrm{HONO}$ and $\lambda_\textrm{LUNO}$, are obtained from the geminal diagonalization. Then, the $\lambda_\textrm{HONO}/|\lambda_\textrm{LUNO}|$ ratio of the RVB wave function is also plotted in Fig.~\ref{acenes:figure:lambda_H2L}. It is apparent that for both JDD and RVB wave functions the ratio decreases as the acene size increases, signaling a stronger multi-radical character for higher acenes. This is in common with many other correlated methods. However, the difference between the JDD and RVB wave functions is striking as far as the magnitude of their ratios is concerned. Indeed, while the JDD ratio extrapolates to 1 (perfect diradical condition) for $n$-acenes with $n > 14$, the RVB ratio is from 3 to 5 times larger in the $n \in [3,9]$ range. Moreover, it is difficult to make any extrapolation out of the RVB ratio, as the corresponding values are quite noisy and very far from the diradical condition. Up to 9 rings - the largest size investigated in this work -, the RVB electronic structure has a much weaker diradical OSS character, compared to JDD.

\begin{figure}[h]
\centering 
\begin{tabular}{c} 
      \includegraphics[width=0.85\columnwidth]{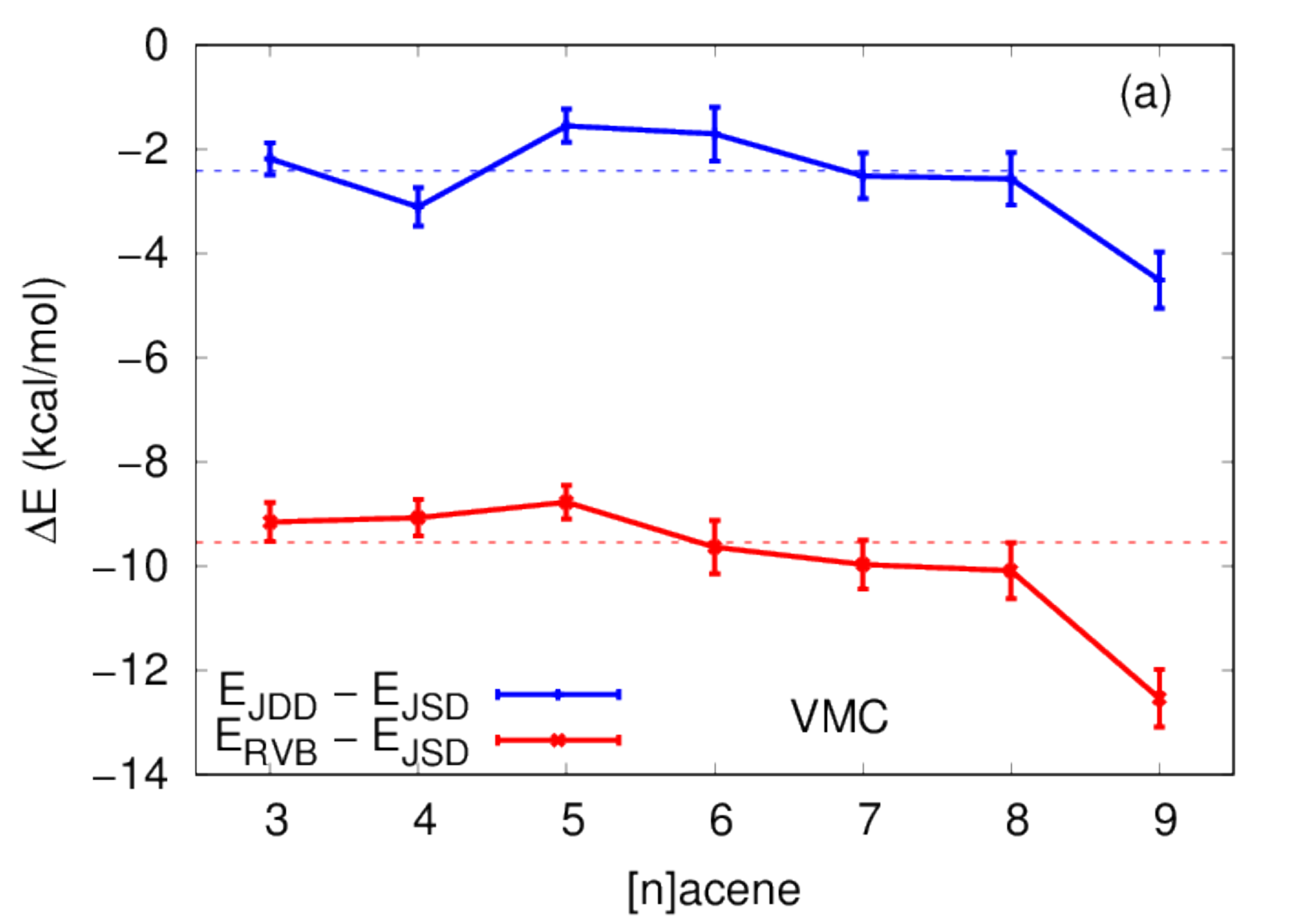} \\
      \includegraphics[width=0.85\columnwidth]{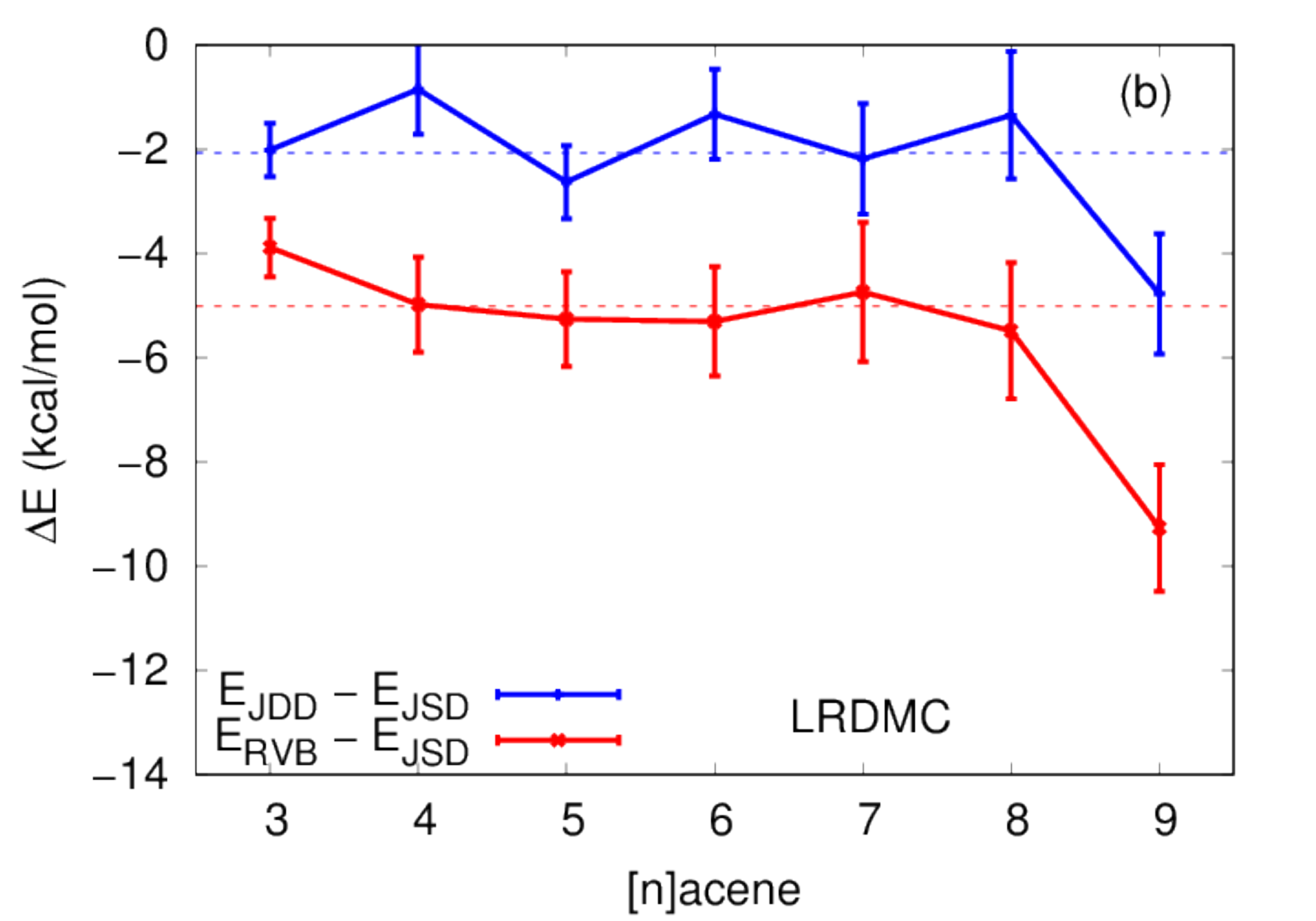} \\
\end{tabular}
\caption{\label{acenes:figure:JAGP_energies}
Energy gain of the JDD and RVB singlets with respect to the best single-reference JSD closed-shell singlet, computed at both VMC (panel (a)) and LRDMC (panel (b)) levels, for the $n$-acenes. It is a measure of the size of static correlation effects on the molecular energetics. Only for the 9-acene, the geometry has been relaxed at the VMC level. For the other molecules, we used the R-B3LYP geometry of the singlet for the JSD calculations, and the B3LYP geometry of the triplet for the JDD and RVB wave functions. The dashed lines are the energy gain averaged over $n \in [3,9]$. 
}
\end{figure}

In order to determine which is the closest wave function to the true ground state of the oligoacenes, we can rely upon the variational principle and compare the variational energies. We compute the energetics of the JSD, JDD and RVB states at both VMC and LRDMC levels. The results as a function of the molecular size are shown in Fig.~\ref{acenes:figure:JAGP_energies}. We see that both JDD and RVB energies are lower than the JSD-CSS reference, and the RVB wave function is systematically better than the JDD at the VMC level. The same hierarchy is confirmed by the most accurate LRDMC method. The energy gain with respect to the CSS solution is sizable, and it seems to depend very weakly on the number of rings, being on average 2 and 5 kcal/mol lower than the JSD reference for the JDD and RVB states, respectively.

From this analysis, we can conclude that the RVB wave function is a better representation of the acenes GS than the JDD one. From our finding, it is clear that the emergent picture is a highly-correlated multi-reference GS, where however the diradical OSS character is significantly weakened, although always present in this class of systems for large enough sizes. In the next Section, we will provide a direct comparison with previous literature and a deeper characterization of the GS physical properties.

We conclude this section on the energetics, by noting that if one defines the condensation energy $\epsilon_\textrm{cond}$ as the difference between the JSD and RVB energies, it seems that $\epsilon_\textrm{cond}$ scales very weakly with the system size. This would imply a vanishing condensation energy per particle in the thermodynamic limit, and thus the absence of any superconducting instability in wires made of zig-zag carbon chains. This is an interesting outcome, as the possible stabilization of a superconducting state has already been proposed in the acenes\cite{kivelson_polyacene_1983}. In that proposal, the superconductivity was phonon-driven, namely triggered by structural distortions and vibrations of the chain. Here, based on our data, we would exclude superconductivity coming from a purely electronic mechanism. This statement must however be confirmed by calculations of larger acenes, in order to have a more precise information on the scaling of $\epsilon_\textrm{cond}$ with respect to the length of the chain.

\subsection{Ground state properties}
\label{acenes:GS}

\begin{figure}[h]
\centering 
\includegraphics[width=1.0\columnwidth]{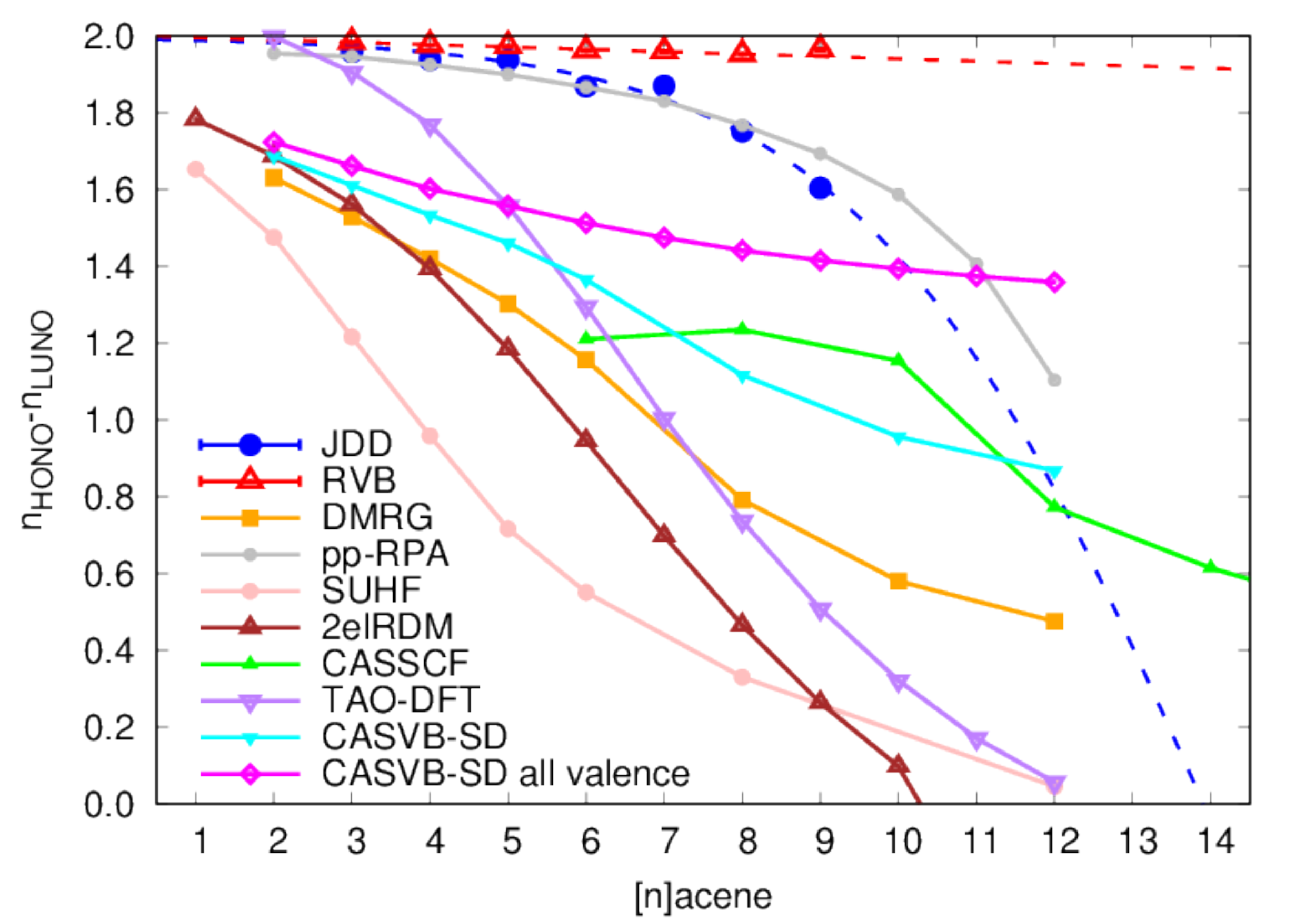}
\caption{HONO-LUNO occupation gap for various methods appeared in literature, plotted together with our JDD (blue points) and RVB (red points) results. We show the values taken from the DMRG\cite{chan_dmrg}, the particle-particle random phase approximation (pp-RPA, Ref.~\onlinecite{yang2016}), the spin-projected unrestricted Hartree-Fock (SUHF, Ref.~\onlinecite{rivero2013}), the two-electron reduced-density-matrix-driven complete active space self-consistent field method (2elRDM, Ref.~\onlinecite{fosso2016}), the complete active space self-consistent approach with H{\"u}ckel-based selection scheme applied to cyclacenes (CASSCF, Ref.~\onlinecite{battaglia2017}), the thermally-assisted-occupation B3LYP (TAO-DFT, Ref.~\onlinecite{chai2017}), the coupled-cluster valence-bond singles and doubles theory restricted to the $\pi$ valence (CASVB-SD, Ref.~\onlinecite{lee2017}), and the one extended to the full valence space (CASVB-SD all valence, Ref.~\onlinecite{lee2017}). The dashed lines are extrapolations of our QMC-based data for JDD and RVB wave functions computed in the $n \in [3,9]$ range. The HONO-LUNO occupation gap closure, found in the extrapolated JDD solution for $n \ge 14$, is compatible with the behavior of the $\lambda_\textrm{HOMO}/|\lambda_\textrm{LUMO}|$ ratio, plotted in Fig.~\ref{acenes:figure:lambda_H2L}. 
\label{acenes:fig:comp_occ_HL_JAGP}  
}
\end{figure}

To compare quantitatively our results with previously published work, in Fig.~\ref{acenes:fig:comp_occ_HL_JAGP} we plot one of the key features to characterize the GS properties, namely the HONO-LUNO occupation gap as a function of the acene size. This is a crucial quantity in order to probe the degree of diradicality of the molecule. A perfect diradical species will have zero gap, while the tendency to diradicality in the large-size limit is signaled by the reduction of the HONO-LUNO occupation gap as the total rings number $n$ increases. 

The HONO-LUNO occupation gap is reported in Fig.~\ref{acenes:fig:comp_occ_HL_JAGP} for various methods. All of them agree qualitatively on the tendency to diradicality for higher acenes.
The difference is in the \emph{magnitude} of this effect. The actual spread among the published methods is remarkable, and increases with the system size. This tells how difficult is to have an accurate quantitative description of the GS of higher oligoacenes. A partial and limited source of difference could come from variations in the definition of the actual plotted quantity, or from differences in the actual system. For instance, the CASSCF calculations of Ref.~\onlinecite{battaglia2017} are for cyclacenes, namely the closed chain version of the oligoacenes. However, the properties of cyclacenes should tend to those of oligoacenes for large enough size. For the particle-particle RPA (ppRPA) calculations\cite{yang2016}, the plotted quantity is the weight of the MO dominant configuration in the RPA expansion. For the thermally-assisted-occupation B3LYP (TAO-B3LYP) approach\cite{chai2017}, the reported thermally populated occupations should tend to the natural orbitals ones by virtue of the chosen effective temperature.
For our QMC calculations, we compute the occupations of the NO representation of the determinantal part, which strictly speaking are not the natural occupations of the full wave function. However, it is the determinant which bears the multi-reference character of diradicality, so we do not expect major differences from the Jastrow factor, which mainly takes into account dynamic correlations.
Except for these caveats, the quantities in Fig.~\ref{acenes:fig:comp_occ_HL_JAGP} should be one-to-one comparable, because they all agree on the following key feature: for a perfect diradical, they all must give a vanishing gap. From this comparison, one can note a common trend. Hartree-Fock has the tendency to overestimate the diradicality, even when the proper symmetry has been restored, such as in the spin-projected unrestricted Hartree-Fock (SUHF) variant\cite{rivero2013}. The TAO-DFT, built on the B3LYP functional, seems to follow an HF-like trend, giving a fast decay of the gap with the number of rings. The slowest decay, corresponding to the least pronounced diradical character, is found for theories able to include a large amount of dynamic correlation, such as the RPA and the CASVB-SD approach, which correlates not only the $\pi$-space but the full set of valence electrons (CASVB-SD all valence)\cite{lee2017}. Our JDD-based QMC result is very much in agreement with RPA, yielding the same behavior of the gap versus $n$.

Our QMC result based on the best variational wave function, i.e. the RVB ansatz, gives an even slower decay than JDD and RPA. By including more resonances in the system, the tendency towards diradicality is strongly reduced. Apparently, a full treatment of correlations acts against an open-shell singlet instability. As mentioned before, this conclusion is drawn from the study of the determinant coefficients, which are optimized together with the Jastrow factor, to minimize the variational energy of the wave function. To better understand this behavior, let us analyze other descriptors, that take into account the combined action of dynamic and static contributions of the fully correlated wave function.

\begin{figure}[h]
\centering 
\includegraphics[width=1.0\columnwidth]{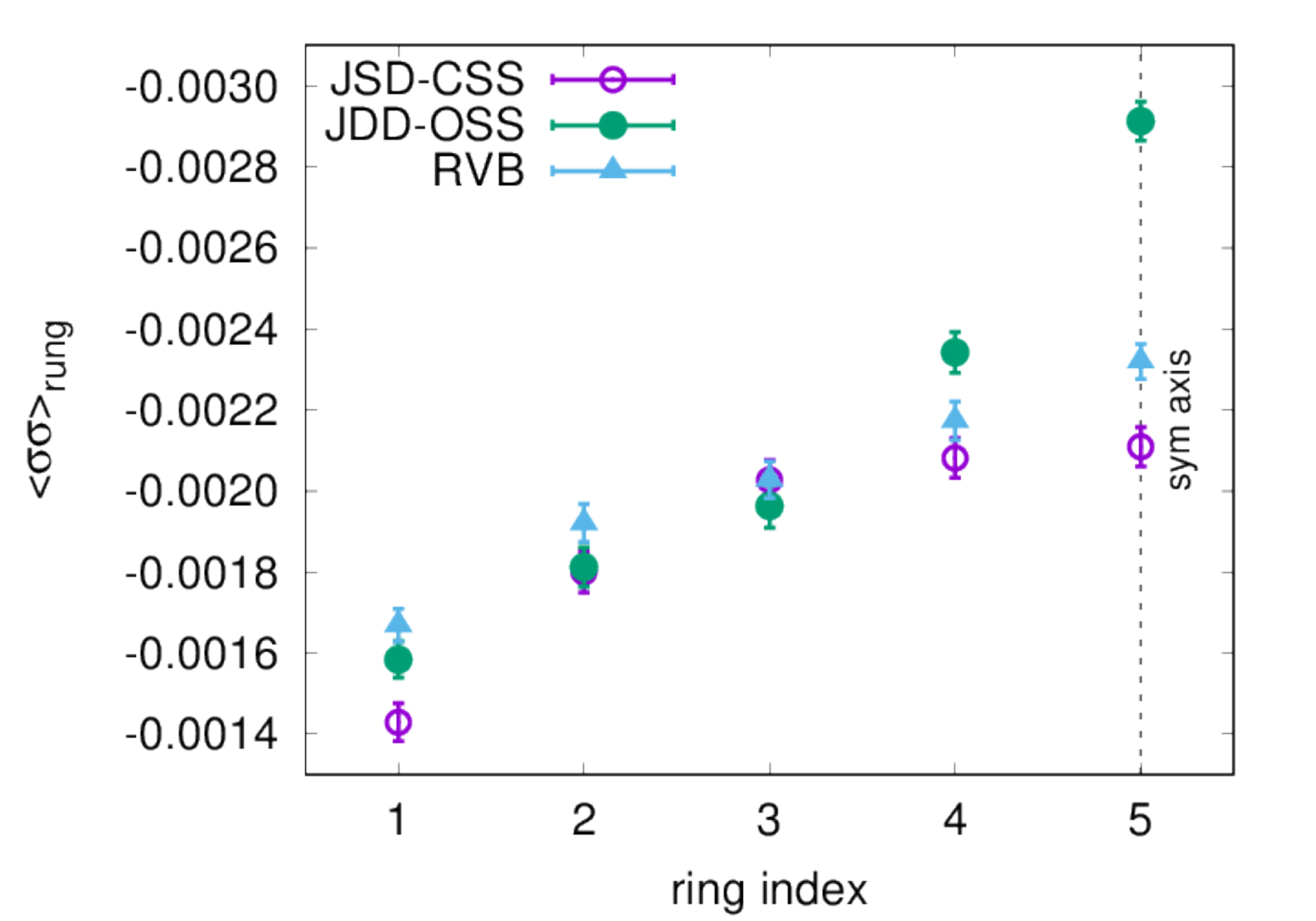}
\caption{Spin-spin correlation functions $\langle \hat{\sigma}_z^i \hat{\sigma}_z^j \rangle_\textrm{rung}$, where $i$ and $j$ belong to rungs involving the most external carbon sites of the nonacene molecule. The corresponding ring is reported in the $x$ axis. We plot the spin-spin correlations for the JSD-CSS (violet open circles), JDD-OSS (green filled circles), and RVB (light-blue filled triangles) states. 
\label{acenes:figure:spinspin} 
}  
\end{figure}

A quantity directly related to the presence of the spin-polarized edge states is the spin-spin correlation function $\langle \hat{\sigma}_z^i \hat{\sigma}_z^j \rangle_\textrm{rung}$ between opposite-edge carbon sites $i$ and $j$.
The $\hat{\sigma}_z^i$ operator counts the electrons belonging to the site $i$, weighted by their $s_z$ spin value. The ``site'' $i$ is defined as the cylinder of radius 1.3$a_0$, whose axis is vertical with respect to the molecular plane and passing through a carbon center $\mathbf{q}_i$. In order to keep the $p_z$ electrons alone within this region,
we exclude from the cylinders a 2.4$a_0$ thick layer containing the molecular plane in the middle. In this way, the ``in-plane'' $sp^2$ electrons do not contribute to the ``site'' correlators. A perfect diradicality involves the localization of $p_z$ electrons into spin-polarized edge states, with an antiferromagnetic correlation between the right and the left edges. Thus, the strength of this correlation is directly related to the diradical character. The result of the spin-spin correlation analysis for the nonacene molecule is plotted in Fig.~\ref{acenes:figure:spinspin}. All three wave functions give a non-negligible antiferromagnetic correlation between the most external rung sites. This increases by going from the external ring to the central one. However, there is a notable difference between the JSD, JDD, and RVB spin-spin correlations in the central ring, where the electron localization is supposed to be the strongest, as revealed by the spin density plotted in Fig.~\ref{acenes:figure:triplet_spin_density} for the triplet state. Indeed, there is a well-defined hierarchy among the three variational models, with the JDD clearly showing the strongest antiferromagnetic correlation. The JSD and RVB are significantly weaker than JDD, with the RVB state being slightly more antiferromagnetic than the JSD. This is in accordance with the HONO-LUNO analysis done before for the determinant only. The JDD has the strongest diradical character, which is weakened when the RVB correlations are included.

\begin{figure}[h]
\centering 
\includegraphics[width=1.0\columnwidth]{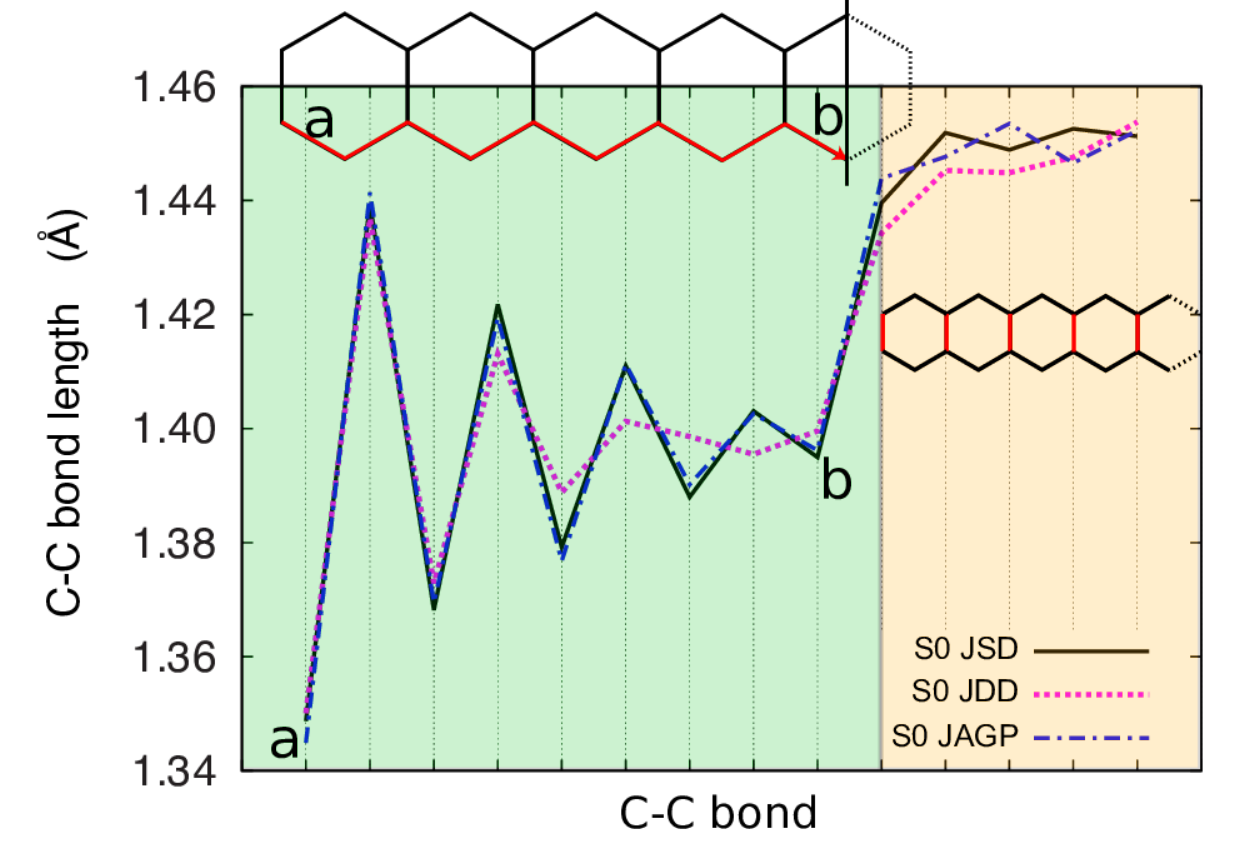}
\caption{Same notations as in Fig.~\ref{acenes:figure:triplet_vs_singlet_geo}, this time for the three singlet states of nonacene taken into account in our work, namely the JSD, JDD and JAGP (RVB) wave functions. All geometries are optimized at the VMC level. 
\label{acenes:figure:bl_S0_QMC}
}  
\end{figure}

To further strengthen our outcome, it is interesting to look at the
equilibrium geometries, relaxed at the VMC level for each variational wave function.
Indeed, as we have argued in
Sec.~\ref{acenes:singlet-triplet_gap}, the bond alternation can reveal the
localization properties of the frontier electrons. As we have already mentioned, we expect that in
the open-shell limit, the unpaired electrons will be localized at the
edge of the molecules, in the same way as the unpaired electrons are localized in the triplet state.
In Fig.~\ref{acenes:figure:bl_S0_QMC}, we plot the
JSD, JDD, and JAGP/RPA geometries, with the same notations as the ones
used in Fig.~\ref{acenes:figure:triplet_vs_singlet_geo}. The JSD has been initiated from the corresponding R-B3LYP geometry of the singlet, while the JDD and JAGP geometries have been started from the triplet. The QMC relaxed geometry of
the JAGP state drifts towards the closed-shell JSD equilibrium geometry,
while the JDD geometry shows an intermediate behavior between the
triplet and the JSD one, as plotted in Fig.~\ref{acenes:figure:triplet_vs_singlet_geo}, reflecting a partial localization of the unpaired
electrons on its edges. This confirms that the JAGP wave function,
although it includes a multiconfigurational character, is not prone to
edge localization, nor to an open-shell arrangement of its electrons, at least up to 9 rings.
The geometry difference, and particularly the difference on the BLA, between the edge-localized (OSS and triplet) and more closed-shells electrons has already been found in Refs.~\onlinecite{jiang2008electronic,qu2009}, with the diradical/polyradical states showing a significant BLA reduction, or a bond length equalization. This strong interplay between electronic structure and geometrical properties found in oligoacenes is very interesting. It is related with the tendency of the infinite-chain system to open the HOMO-LUMO gap by gaining internal energy, and with the quest of the most effective way to open such a gap. One of these ways is the stabilization of the edge-localized antiferromagnetic order implied by the OSS, the other way is to open a gap by breaking the spatial symmetry of the benzenoid ring, such as the appearance of a sizable BLA. Our QMC geometries and spin-spin correlation functions suggest that there is a tight competition between these two phenomena. The JAGP wave function points at the latter scenario, while the JDD wave function at the former. As we know from their variational energies, the winner is the JAGP wave function over the JDD one. More generally speaking, this result shows the importance of optimizing the geometry in the QMC correlated framework.

\begin{figure*}[t!]
\centering 
\begin{tabular}{ccc} 
\includegraphics[width=0.66\columnwidth]{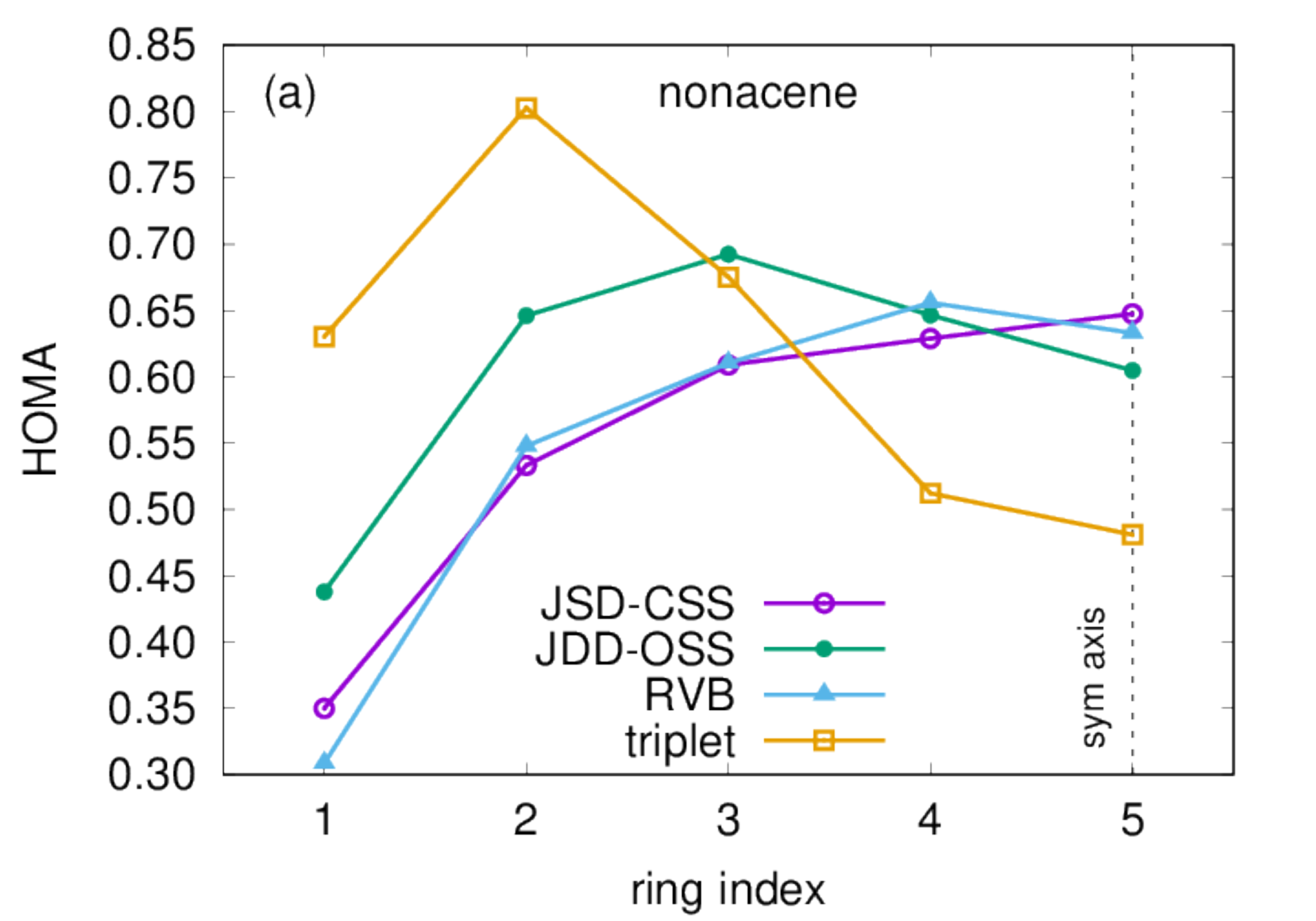}&
\includegraphics[width=0.66\columnwidth]{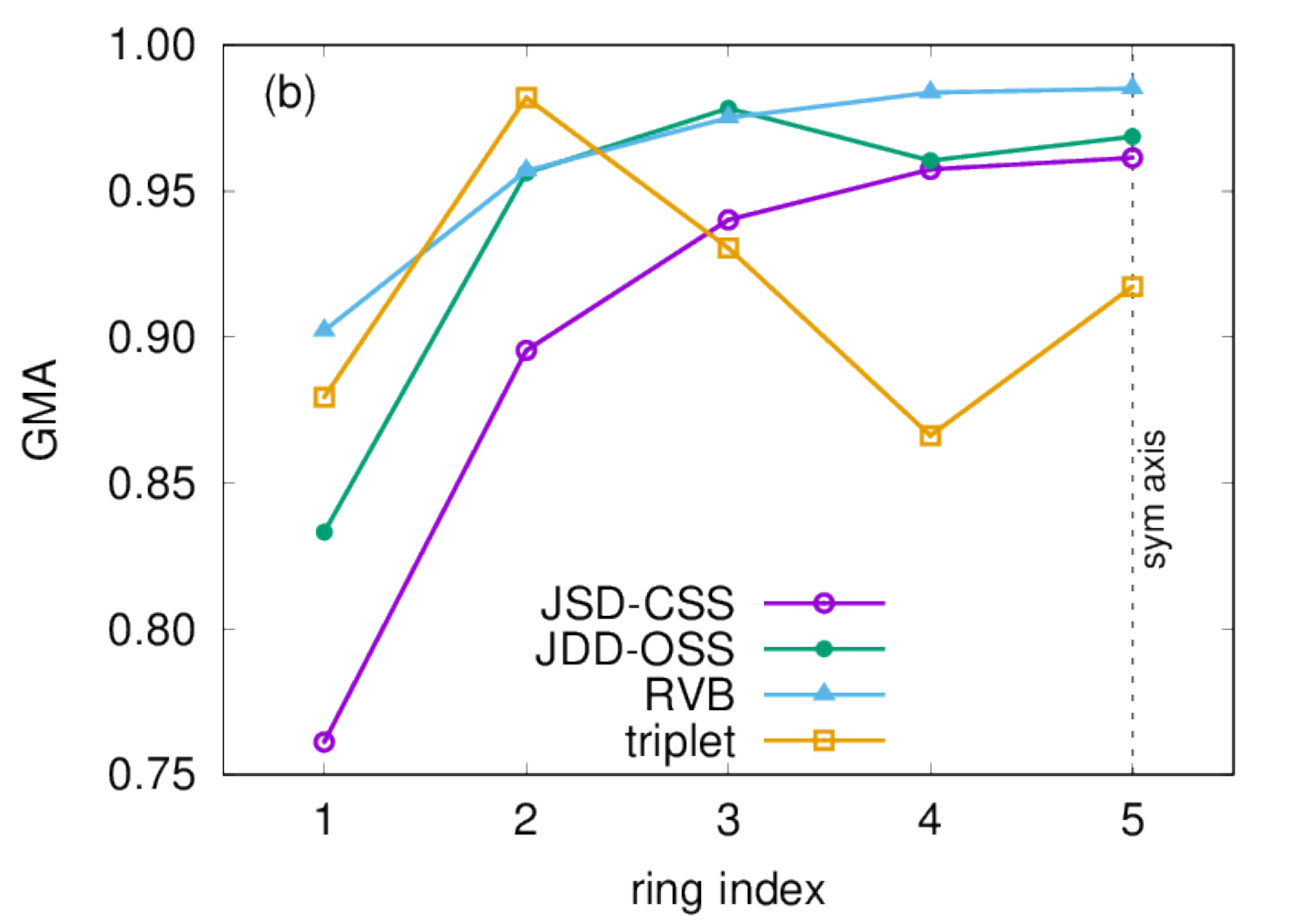}&
\includegraphics[width=0.66\columnwidth]{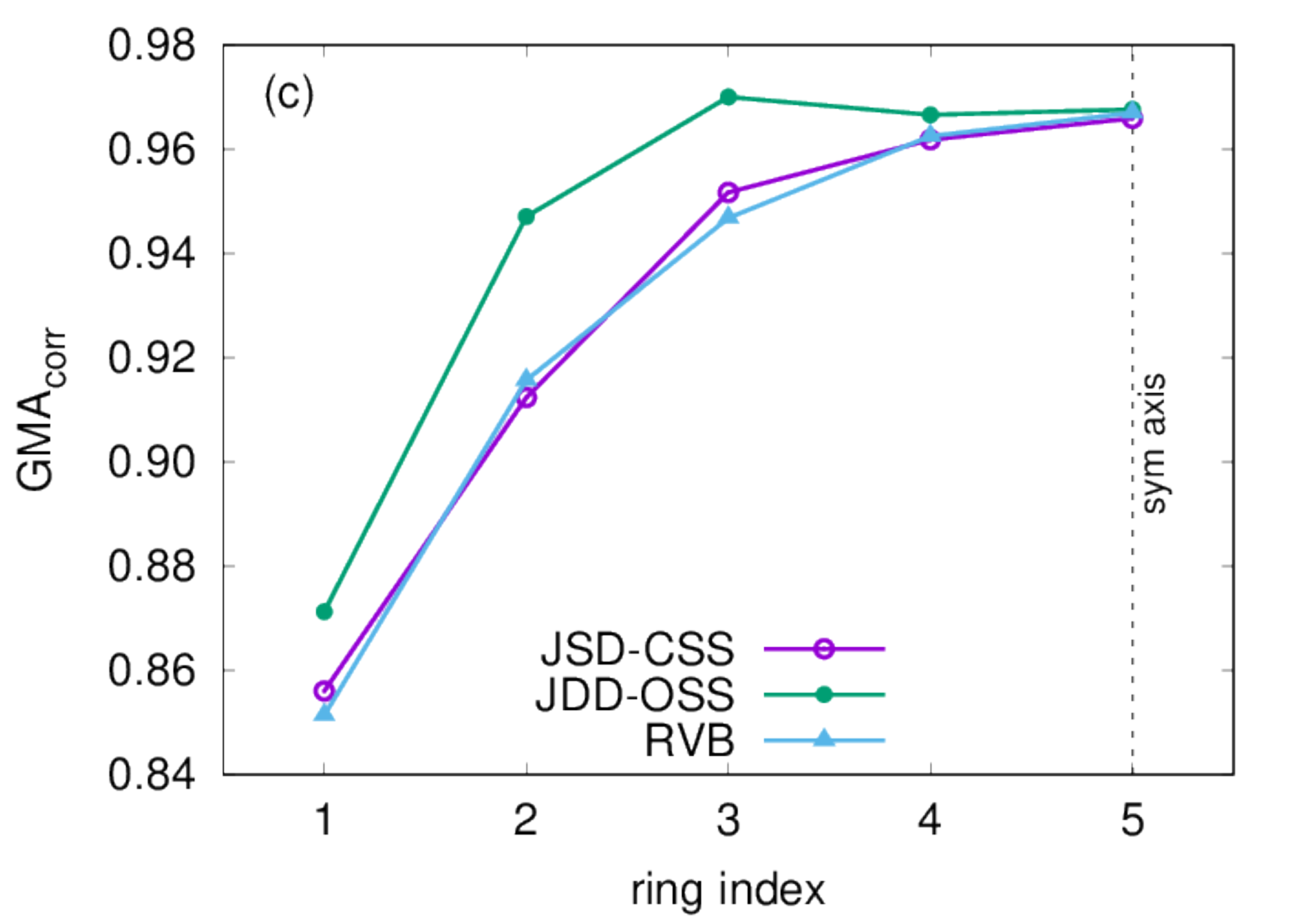} \\
\end{tabular} 
\caption{Panel (a): HOMA aromaticity index (Eq. ~\ref{HOMA}) calculated for each fused ring in the nonacene molecule, based on the relaxed geometry of the JSD-CSS, JDD-OSS, RVB, and triplet wave functions. The ring index goes from the outermost ring (1) to the central one (5), following the same order as the one used to plot the bond lengths of Figs.~\ref{acenes:figure:triplet_vs_singlet_geo} and \ref{acenes:figure:bl_S0_QMC}. Panel (b): geminal measure of aromaticity (GMA), as defined in Eq.~\ref{GMA}), with the same notations as in panel (a). Panel (c): same as panel (b), but for the correlated geminal measure of aromaticity (GMA$_\textrm{corr}$). 
\label{acenes:figure:HOMA}
}  
\end{figure*}

We argued that the geometry variation between RVB, JDD, and triplet wave functions is so strong because of the direct consequences of electron localization in destabilizing the Kekul\'e resonances, which instead have the tendency to delocalize the electrons around the benzenoid ring. In order to be more quantitative in this respect, we computed the harmonic oscillator measure of aromaticity (HOMA) index\cite{kruszewski1972}, defined as
\begin{equation}
\textrm{HOMA}=1- \frac{\alpha}{6} \sum_i^6  (d_i - R_\textrm{opt})^2,
\label{HOMA}
\end{equation}
with $d_i$ the length of the $i$-th C-C bond in the benzenoid ring, and $\alpha$ and $R_\textrm{opt}$ parameters. We used $\alpha=257.7$ and $R_\textrm{opt}=1.388 \AA$, calibrated on the aromaticity level appropriate for our system\cite{krygowski1993}. The unitary value corresponds to the ideal case of full resonance. The HOMA indices are plotted in Fig.~\ref{acenes:figure:HOMA}(a) for our 4 wave functions, as a function of the ring positions in the nonacene molecule.
Surprisingly, the BLA along the zig-zag bonds of the oligoacenes does not necessarily imply the weakest aromaticity index. In fact, the JDD-OSS, which shows a bond equalization in the central rings, has a weaker aromaticity than JSD and JAGP, exactly for the same rings. This is even more apparent for the triplet state, which has electrons sharply localized at the edges, as we have seen in Sec.~\ref{acenes:singlet-triplet_gap}. Indeed, a key role is played by the central rung bond, which is elongated in the case of stronger diradicality, as drawn in the right panel of Figs.~\ref{acenes:figure:triplet_vs_singlet_geo}, because the diradical pair is responsible for a weaker non-neighbor bond of Dewar type crossing the central ring. Moreover, both JDD-OSS and triplet states share the same HOMA behavior as a function of the ring position. They have a \emph{reduced} aromaticity in the central ring, where the Kekul\'e resonances are fragilized by the electron localization. This is a clearcut signature that the JDD-OSS state follows the same physics as the triplet state, as far as the electron localization is concerned. Moreover, this is in contrast with the behavior of JSD and JAGP geometries, which have instead an \emph{increased} aromaticity in the center of the molecule, signature of more delocalized electrons, as reflected by a weaker diradical character. This relation between aromaticity and diradicality has already been found in Ref.~\onlinecite{poater2003}, calculated on relaxed R-B3LYP and U-B3LYP geometries, where the OSS was a spin-contaminated symmetry-broken solution. Our analysis strengthens the previous outcome, as it is based on explicitly correlated non spin-contaminated wave functions. The main finding here is that the RVB, our best and most correlated wave function, behaves much more as a CSS. The HOMA index descriptor confirms therefore that by including a full set of RVB resonances in the JAGP wave function, a stronger aromaticity is recovered in the central ring, with less diradical character. The full treatment of dynamic and static correlations acts against a diradical instability.

The HOMA is an aromatic index based on geometric considerations. Nevertheless, the aromaticity level is sensitive to the quantity used to probe it\cite{balaban2005,fowler2011,radenkovic2017}. Indeed, aromaticity by itself is not a direct observable. For this reason, it is useful to double check the outcome of the HOMA index by means of a complementary probe. Possibly, we would also like to quantify the aromaticity at the electronic level. With this aim, we exploit the picture provided by the AGP wave function. Indeed, the $\lambda_{i,j}$ matrix of the geminal $\phi$ expanded in the minimal basis set, yields directly the valence bond strength between the sites $i$ and $j$. It is clear that in the case of a perfect resonance (full aromaticity), all VB amplitudes around a ring will be equal, while in the case of a dimerized structure (zero aromaticity), they will strongly alternate, vanishing where the covalent bond is absent. Thus, we introduce a new descriptor that reveals these properties, dubbed as ``geminal measure of aromaticity'' (GMA). It is defined as follows:
\begin{eqnarray}
\textrm{GMA} & = &\frac{\langle \tilde{\lambda}_i \tilde{\lambda}_{i+1} \rangle}{\langle \tilde{\lambda} \rangle^2}, \nonumber \\
\textrm{with~~~} \langle \tilde{\lambda} \rangle & = & \frac{1}{6} \sum_i^6 \tilde{\lambda}_i, \nonumber \\
\textrm{and~~~} \langle \tilde{\lambda}_i \tilde{\lambda}_{i+1} \rangle & = & \frac{1}{6} \sum_i^6 \tilde{\lambda}_i\tilde{\lambda}_{i+1} ,
\label{GMA}
\end{eqnarray}
where $i$ is the bond index running over a given benzenoid ring in a spatial-sequential order, and $\tilde{\lambda_i} = |\lambda_{l,m}|$ is the absolute value of the valence bond strength between sites $l$ and $m$, indexed by $i$. Only $\lambda_{l,m}$ connecting $p_z$ orbitals are taken into account in Eq.~\ref{GMA}.  It is easy to show that $0 \le \textrm{GMA} \le 1$, with $\textrm{GMA}=0$ for a ``perfectly dimerized'' structure, and $\textrm{GMA}=1$ for a ``perfect'' aromatic structure. To get $\phi$ in the minimal basis set, we project the AGP function fully optimized in the extended JAGP framework onto the minimal basis, optimally contracted by means of the geminal embedding scheme\cite{sorella2015}. The resulting GMA index is plotted in Fig.~\ref{acenes:figure:HOMA}(b). The GMA is in qualitative agreement with the HOMA. The aromaticity increases for both the RVB and JSD wave functions, as in the HOMA, while the JDD follows qualitatively the behavior of the triplet wave function, with a minimum of aromaticity in the fourth ring (it is the central 5-th ring in the HOMA), and a maximum located as in the HOMA. 

As in the case of the HONO-LUNO gap analysis, the GMA in Eq.~\ref{GMA} involves only the AGP $\lambda_{l,m}$ matrix. One would like to explicitly include the effect of the Jastrow factor in the aromaticity measure. To do so, one has to evaluate the valence bond strength $\tilde{\lambda_i}$ for the correlated wave function. A natural extension of the $\tilde{\lambda_i}$ definition is based on the fact that the valence bonds are made of \emph{singlets}, therefore their strength is given by the dimer correlation between two singly-occupied sites. One can thus write $\tilde{\lambda_i} = -\langle  \hat{\sigma}_z^l \hat{\sigma}_z^m \rangle$, where the definition of ``site'' has been provided before. We will call GMA$_\textrm{corr}$ the resulting measure of aromaticity, which is plotted in Fig. ~\ref{acenes:figure:HOMA}(c) for our three nonacene spin-singlet states. It is apparent that the full inclusion of correlation in the GMA$_\textrm{corr}$ does not change the qualitative behavior already provided by both HOMA and GMA, with the main feature being the reduction (increase) of aromaticity for the JDD (JSD/JAGP) wave functions when approaching the central ring. However, the explicit inclusion of the Jastrow factor in the VB strength estimates brings the JSD and RVB aromaticities close to each other, as in the HOMA, while in the simpler GMA the aromaticity of the RVB wave function turns out to be always higher. This points to the importance of the interplay between the Jastrow and the AGP part in setting the ultimate properties of the system, as highlighted for instance in Refs.~\onlinecite{neuscamman2012size,neuscamman2016,goetz2017}.

\begin{figure}[h]
\centering 
\begin{tabular}{c} 
\includegraphics[width=0.85\columnwidth]{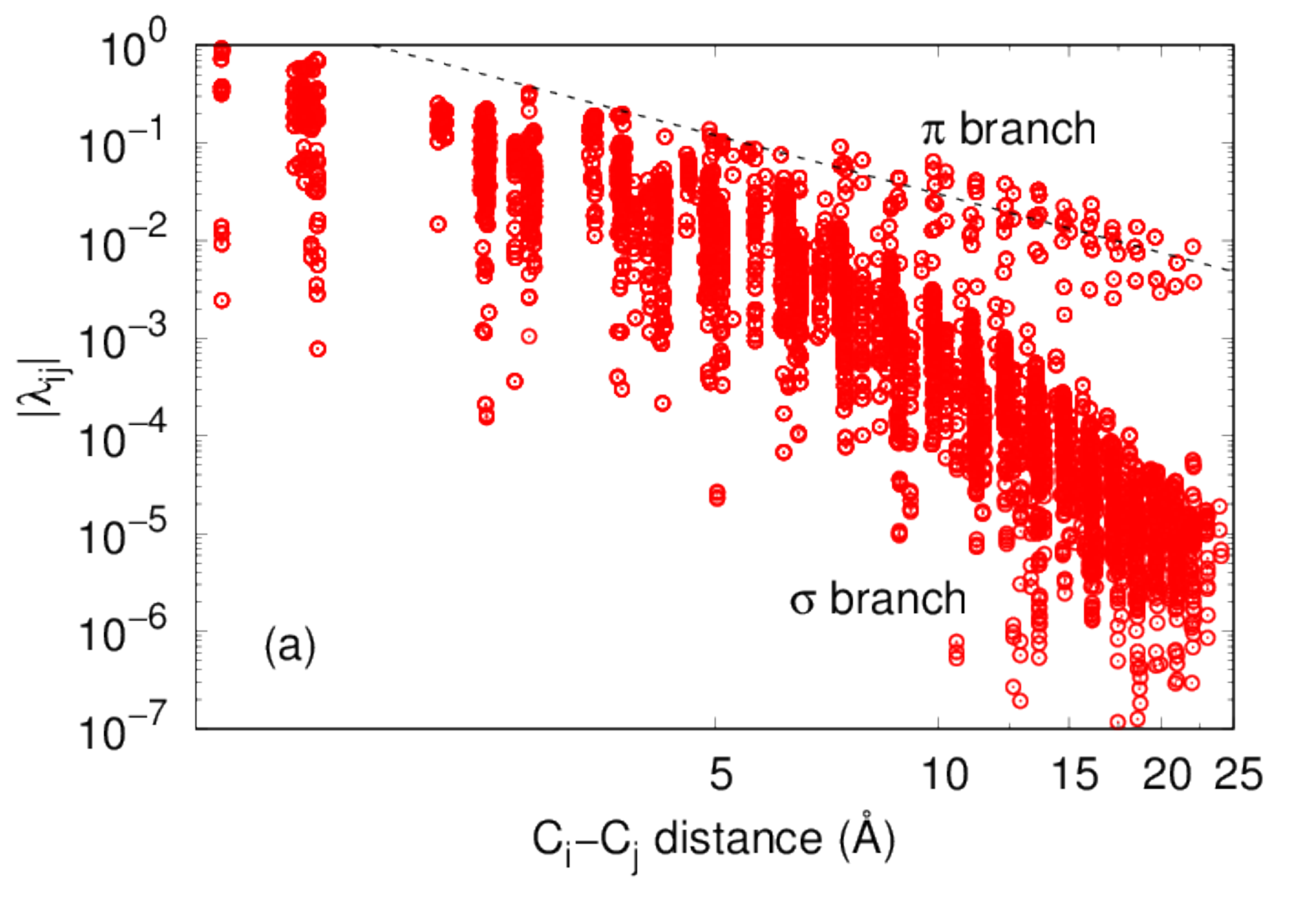} \\
\includegraphics[width=0.85\columnwidth]{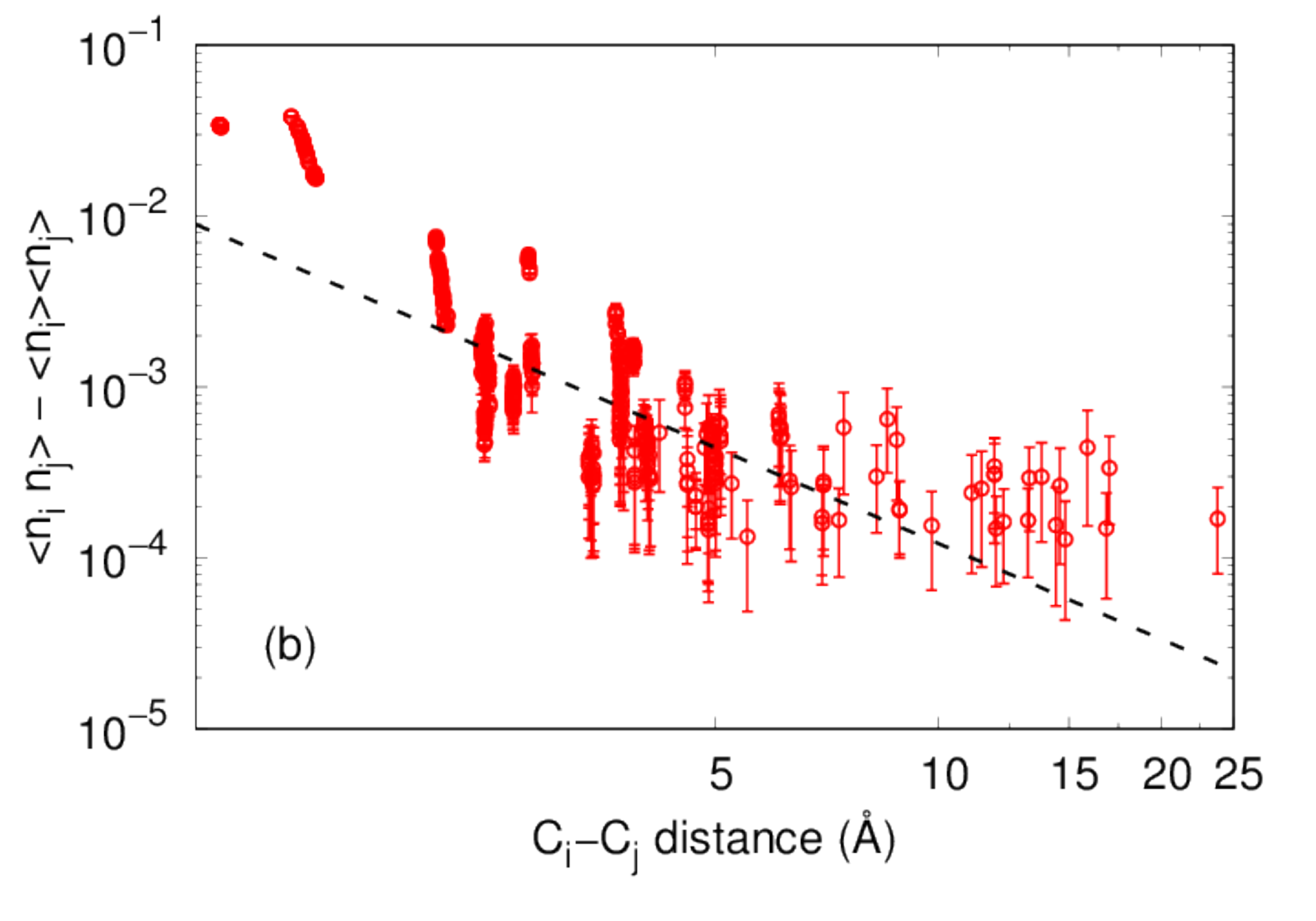}\\ 
\end{tabular} 
\caption{Panel (a): log-log plot of the $\lambda_{i,j}$ geminal coefficients linking the atomic orbitals of $i$ and $j$ carbon sites, as a function of the C$_i$-C$_j$ intersite distance $d$. The $\lambda_{i,j}$ coefficients are obtained by projecting the fully optimized RVB wave function into a minimal basis set. The $\pi$ branch connecting $p_z$ orbitals, and the $\sigma$ one, involving the $sp^2$ hybridized orbitals, are clearly separated, owing to their very different decay with the C-C distance (power law versus exponential). The dashed line is the power-law decay $1/d^\alpha$ with exponent $\alpha=2$. Panel (b): log-log plot of the charge-charge correlation function $\langle \hat{n}_i \hat{n}_j \rangle - \langle \hat{n}_i \rangle \langle \hat{n}_j \rangle$ computed for the many-body RVB wave function. From the definition of ``site'' which enters the many-body correlator (see text), only the $\pi$ electrons are involved in the measured charge fluctuations. The dashed line is a least-square fitting of the function $\delta/d^\gamma$. The best fit is found for $\gamma=1.9 \pm 0.3$. 
\label{acenes:fig:lambda_det}
}
\end{figure}

The HOMA, GMA, and GMA$_\textrm{corr}$ are ``local'' probes, as they depend on the nearest-neighbors only. We conclude this section with an analysis on the long-range properties of the valence bonds strength in the oligoacenes. As usual, we take the nonacene as our favorite test case. The minimal-basis $\lambda_{i,j}$ are shown in a log-log plot in Fig.~\ref{acenes:fig:lambda_det}(a), as a function of the C$_i$-C$_j$ intersite distance $d$ of the corresponding valence bond. The decay of the in-plane $\sigma$ bonds is exponential, while
the $\pi$ bonds are much more slowly decaying, in a way that is compatible with a $1/d^\alpha$ power-law behavior with exponent $\alpha \approx 2$. The long-range nature of the $\pi$ bonds reveals that a highly entangled ground state can be formed, once all resonances are taken into account. This goes beyond the ``classical'' picture of the Kekul\'e diagrams, where only the nearest-neighbor configurations are usually drawn. Having $\phi$ the same eigenvectors as the AGP one-body reduced density matrix (1-RDM), it shares with it also its spatial decay. It is well known that the 1-RDM localization is related to the conductivity of the system\cite{kohn1964,gallo2016}. The $1/d^2$ behavior found in Fig.~\ref{acenes:fig:lambda_det}(a) is a signature of a metallic or quasi-metallic character\cite{goedecker1998}. Indeed, the Jastrow function $J$ decays as $1/d^\beta$ with $\beta \ge 1$, thus at large distance it is not sufficiently strong to freeze the holons-doublons charge fluctuations allowed by the AGP part\cite{capello_superfluid_2007,capello_mott_2008}, and open a charge gap. If this behavior is preserved in the infinite-chain limit, the polyacene will clearly be a metal.

The quasi-metallic properties are also confirmed by the QMC calculation of the charge-charge correlation function $\langle \hat{n}_i \hat{n}_j \rangle - \langle \hat{n}_i \rangle \langle \hat{n}_j \rangle$ between sites $i$ and $j$, where the site definition is the same as in the spin-spin correlation function, introduced previously in the paper. The operator $\hat{n}_i$ counts the number of electrons falling in the site $i$. These charge-charge correlations are computed for the JAGP wave function, and plotted as a function of the intersite distance in Fig.~\ref{acenes:fig:lambda_det}(b). As expected from the above analysis, we find the same decay as the one of the AGP lambda matrix, namely a power-law behavior with exponent close to 2. This decay is compatible with a metallic character for quasi-1D systems\cite{capello2005,casula2006}. A similar analysis can be carried out for the spin-spin correlation functions. In the same way, one can show that their long-range decay is a power law with exponent $\approx 2$. Thus, also the spin spectrum will be gapless in the the infinite-chain limit.

\section{Conclusions and perspectives}
\label{acenes:conclu}

In this paper, we have studied the ground state properties of the $n$-acenes, with $n \in [3,9]$. By means of QMC techniques, we have employed a variational JAGP wave function that is the \emph{first-principles} representation of the RVB ansatz, the ideal many-body framework to study polycyclic aromatic hydrocarbons. Resolving the ground state of oligoacenes is a very challenging task, due to the strongly correlated nature of these one-dimensional (1D) systems. A precise description of both static and dynamic correlation effects is needed to meet that goal. A large variety of quantum chemistry methods has been used to tackle this problem, with results that agree only qualitatively on the multi-reference character of the ground state and on its polyradical nature, which increases with the system size. Our RVB wave function includes all necessary ingredients to successfully capture the elusive nature of this ground state. Supplementing the fully optimized variational JAGP wave function with the projective lattice regularized diffusion Monte Carlo method, highly accurate results can be obtained for this class of systems.

Our results are in contrast with the common belief, based on previous data published in literature, that $n$-acenes are strongly diradical already for $n = 6$. We found instead that the RVB ground state has a weak diradical character at least until $n=9$, the largest size we have explored. This is revealed by a number of descriptors, such as the $\lambda_\textrm{HONO}/\lambda_\textrm{LUNO}$ ratio, the spin-spin correlation functions between the two edges, the BLA, and finally the aromatic HOMA and GMA indices. This outcome is substantiated by a direct comparison with another wave function ansatz, the JDD, tailored to describe a perfect diradical. The JDD wave function shows, indeed, a much stronger diradical character, comparable to what found very recently in pp-RPA, and CASVB-SD calculations. It turns out that the RVB variational energy is systematically lower than the JDD one, at both VMC and LRDMC levels, and therefore closer to the true GS of the system. This suggests that a complete treatment of dynamic and static correlations is detrimental to a diradical instability in this class of systems. In any case, the RVB results agree with the common trend of an increased diradicality as a function of the acene length, although with a small slope. This is due to the entangled multi-reference nature of the RVB wave function, which allows for the resonance of an exponentially growing valence bond states. The importance of the valence bonds resonance is highlighted by the long spatial range of the $\lambda_{i,j}$ couplings, which disproves the validity of short-range RVB models. The absence of a diradical instability, and the associated lack of spin-polarized zig-zag borders antiferromagnetically ordered,
is in agreement with the behavior expected from enhanced quantum fluctuations in 1D systems\cite{mermin1966,yazyev2008}. It is also in agreement with the conclusions of Deleuze and coworkers, who showed, based on mathematical arguments, that the ground state of this class of systems cannot have symmetry-broken spin-polarized edges\cite{huzak2011,deleuze2013}. This property is robust against relativistic spin-orbit effects\cite{perumal2012}.

An interesting perspective of this work is to study whether that situation persists in a quasi-1D regime, namely in wider graphene nanoribbons. Although antiferromagnetic edge correlations have been found in DFT calculations of graphene nanoribbons, it is not clear whether RVB correlations will be strong enough to melt the spin order or to prevent electron edge-localization, even for widths larger than a single ring. QMC is certainly a very suitable technique to address these questions on wider ribbons as well. This could be an interesting follow-up of the present study.

\section*{Acknowledgments}
We acknowledge useful discussions with F. Mauri, S. Sorella and B. Bra\"ida.
The computational resources have been provided by the PRACE project number PRA143322, the
GENCI project number 96493, and the RIKEN Advanced Institute for Computational
Science through the HPCI System Research project number G16026, allocated on the HOKUSAI GreatWave HPC facility.

\bibliographystyle{aipnum4-1}
\bibliography{intro,higher_acenes,pioud}

\end{document}